\documentclass[journal]{IEEEtran}
\usepackage{amssymb}
\usepackage[cmex10]{amsmath}
\usepackage{stfloats}
\usepackage{float}
\usepackage{graphicx}
\usepackage{subfigure}
\usepackage{tabularx}
\usepackage{epsfig,epsf,color,balance,cite}
\usepackage{verbatim}
\usepackage{url}
\usepackage{bm}
\usepackage{amsthm}
\usepackage{titlesec}
\usepackage{array}
\usepackage{booktabs} % For better table formatting
\usepackage{makecell} % Allows line breaks in table cells
\usepackage{multirow} % Supports multirow cells
\usepackage{algorithm}
\usepackage{algorithmic}

\usepackage{color}
\definecolor{myc1}{rgb}{0,0,1}

\graphicspath{{./figures/}}

\begin{document}
	
	% paper title
\title{Split and Aggregation Learning for Foundation Models Over Mobile Embodied AI Network (MEAN): A Comprehensive Survey}
% \title{Split and Aggregation Learning for Foundation Models in 6G Communication Systems: A Comprehensive Survey}
%\title{Large AI Model Empowered Integrated Learning and Communication Systems: Design, Opportunities, and Challenges}

\author{Qianzhou Chen,
        Siqi Sun,
        Minrui Xu,
        Sijie Ji,
        Jiawen Kang,~\IEEEmembership{Senior Member,~IEEE,}\\
        Yijie Mao,~\IEEEmembership{Senior Member,~IEEE,}
        Zhouxiang Zhao,
        Zhaohui Yang,
        and Dusit Niyato,~\IEEEmembership{Fellow,~IEEE}

% \thanks{This work is supported by . \textit{(Corresponding author: Zhaohui Yang)}.}
% \thanks{Zhouxiang Zhao and Zhaohui Yang are with the College of Information Science and Electronic Engineering, Zhejiang University, and also with Zhejiang Provincial Key Laboratory of Info. Proc., Commun. \& Netw. (IPCAN), Hangzhou 310027, China (e-mails: \{zhouxiangzhao, yang\_zhaohui\}@zju.edu.cn).}
% \vspace{-1em}
}
		
\maketitle
% \vspace{-0.25em}
\begin{abstract}
 The rapid advancements in foundation models and sixth-generation (6G) wireless communication systems necessitate the development of efficient, scalable, and privacy-preserving machine learning approaches. For foundation models in 6G, split learning (SL) and aggregation learning (AL) have emerged as promising paradigms that address key challenges in distributed artificial intelligence (AI), such as communication efficiency, resource allocation, and data privacy. SL enables multiple entities to collaboratively train deep learning models by partitioning neural networks, while AL focuses on aggregating intermediate results or model updates from multiple participants, improving robustness, optimizing resource utilization, and mitigating data leakage risks. Specifically, SL is ideal for scenarios requiring strict data isolation (e.g., vertical collaborations), whereas AL suits homogeneous horizontal data settings; they can be combined to balance privacy and communication efficiency. This survey provides a comprehensive analysis of SL and AL in 6G communication systems, exploring their architectures, technical methodologies, and integration with AI-native 6G communication technologies. We examine different SL configurations, aggregation techniques, and their roles in optimizing distributed foundation models. Furthermore, we discuss their applications in emerging wireless networks, including semantic communication, reconfigurable intelligent surfaces (RIS), space-air-ground integrated networks (SAGINs), and quantum communication. By analyzing the impact of SL and AL, this survey provides insights into their role in shaping distributed AI-driven communication systems in the 6G era, focusing on efficiency, privacy preservation, and scalability.
\end{abstract}
\IEEEpeerreviewmaketitle
	    	
\section{Introduction}
\subsection{Background and Motivation}

Foundation models, such as large language models (LLMs), have advanced rapidly, demonstrating remarkable capabilities in performing complex tasks across multiple domains. Foundation models have garnered significant attention due to their ability to understand and generate human-like text with remarkable accuracy. Foundation models, while powerful, come with substantial challenges, including the need for enormous amounts of data, computational resources, and, critically, the need to maintain data privacy and security. Traditionally, training these foundation models requires centralized datasets, where data and local models from various sources is aggregated in a single location for training. However, in domains such as healthcare, finance, and communication systems, where data is sensitive and distributed across multiple locations, this approach poses significant risks. As foundation models continue to grow in size and complexity, the need for efficient and secure training methodologies becomes increasingly critical~\cite{zheng2024safely}.

To address these challenges, distributed learning, such as federated learning (FL), split learning (SL), and aggregation learning (AL), has emerged as a solution by distributing the training workload across multiple participants, enabling collaborative training while preserving data privacy. This is particularly relevant in modern wireless communication systems, where data is inherently distributed, and privacy considerations are paramount. FL has been one of the most widely adopted distributed learning paradigms, enabling decentralized training by allowing multiple devices or institutions to train a global model collaboratively without sharing their local data. Instead, only model updates, such as gradients or weights, are shared with a central server, which aggregates these updates to improve the global model~\cite{mcmahan2017communication,le2024applications,jia2025comp}. While FL effectively mitigates some privacy risks, it is not without limitations.

One of the key challenges in FL is the potential for data leakage through model updates \cite{9264742}. Even without sharing raw data, adversaries may infer sensitive information by analyzing the model parameters exchanged during training. Furthermore, FL assumes that all participating devices have sufficient computational resources to train the full model, which may not be the case in practice. This limitation is particularly pronounced in foundation models, where the computational demands are substantial~\cite{li2020federated}. Additionally, FL requires frequent model updates between devices and the central server, leading to high communication overhead. In wireless communication networks, where bandwidth is limited and network conditions fluctuate, this excessive communication burden can degrade both model performance and network efficiency.

As the demand for privacy-preserving and resource-efficient artificial intelligence (AI) training methods grows, a new paradigm known as SL and AL has emerged. SL enables multiple parties to collaboratively train a deep learning model by dividing the model architecture across different entities~\cite{gupta2018distributed}. Each participant is responsible for training a portion of the model, typically the initial or intermediate layers, while the remaining layers are trained by other participants or a central server. This division allows each participant to share only intermediate activations rather than raw data or full model updates, thereby enhancing privacy and security~\cite{vepakomma2018split}. This method is particularly advantageous when dealing with sensitive data distributed across multiple locations, such as in healthcare, finance, or communication systems.

Building on this concept, AL introduces a mechanism where intermediate results from multiple participants are aggregated at certain points in the model training process~\cite{shokri2015privacy}. This aggregation can occur either at the end of a training epoch or dynamically throughout the training process, depending on the specific application and network constraints. By aggregating intermediate activations or model updates, this approach can enhance the robustness of the training process, reduce the impact of outliers, and improve the overall accuracy of the model. The combination of split and aggregation learning offers several key advantages over traditional FL, especially when applied to foundation models.

Wireless communication plays a critical role in enabling SL and AL by providing the infrastructure necessary for distributed training between devices, edge servers, and cloud platforms \cite{10024766,10233741,yang2019energy}. The integration of SL and AL into wireless communication systems, particularly in the era of 6G, presents both opportunities and challenges. On the one hand, 6G networks will offer ultralow latency, high bandwidth, and AI native architectures, which can significantly improve the efficiency of distributed AI training~\cite{letaief2019roadmap}. On the other hand, wireless networks introduce additional constraints such as communication overhead, synchronization issues, and security vulnerabilities. Addressing these challenges requires innovative solutions in communication protocols, resource management, synchronization strategies, and security measures. 

The emergence of 6G wireless communication systems provides an ideal infrastructure for deploying SL and AL at scale \cite{chen2020joint,9562487}. 6G networks are expected to introduce key technological advancements directly benefiting distributed AI, including ultra-reliable low-latency communications (URLLC), intelligent network slicing~\cite{rafique2025survey}, and edge computing~\cite{yang2025beyond}. These capabilities will enable the dynamic allocation of network resources based on AI workload demands, ensuring efficient and adaptive model training \cite{11381448,10577141,yang2025privacy}. Furthermore, advances in semantic communication and reconfigurable intelligent surfaces (RIS) will allow more efficient transmission of model updates, further optimizing communication efficiency~\cite{9955525,yang2023energy,e26050394,11250771,zhang2021envisioning,9410457,10550151}.

\subsection{Related Work}

\begin{table*}
    \centering
    \renewcommand{\arraystretch}{1.5} % 设置行高为默认的1.5倍
    \caption{Related Existing Surveys on SL and AL for Wireless Communications.}
    \begin{tabular}{|p{1cm}|p{4cm}|p{11cm}|}
         \hline
         \textbf{Paper} & \textbf{Topic} & \textbf{Key contribution} \\
         \hline
        ~\cite{qian2022distributed} & Distributed learning for wireless communications & Introduced the typical frameworks and algorithms for distributed learning. Provided examples of applications in different network layers. \\
         \hline
        ~\cite{chen2021distributed} & Distributed learning for wireless communications & Presented an overview of several emerging distributed learning paradigms, including federated learning. \\
         \hline
        ~\cite{hu2021distributed} & Distributed learning for wireless communications & Discussed techniques and applications of distributed learning. Described the potential adversarial attacks and corresponding countermeasures, and summarized open issues. \\
         \hline
        ~\cite{letaief2021edge} & Edge AI & Introduced edge AI system for 6G, including the new design principles, resource allocation, and system architecture.\\
         \hline
        ~\cite{shi2020communication} & Edge AI & Discussed communication-efficient edge AI techniques, from both algorithmic and system perspectives for training and inference tasks at the network edge. \\
         \hline
        ~\cite{chen2024big} & Foundation model & Introduced big AI models for 6G wireless networks. Discussed the opportunities, the challenges, and the potential research directions. \\
         \hline
        ~\cite{duan2022combined} & Federated and split learning & Analyzed combined federated and split learning in edge computing. Identified some open problems and discussed possible directions for future research. \\
         \hline
        ~\cite{thapa2021advancements} & Federated and split learning & Discussed advancements in federated learning towards privacy preservation. Explored the evolution from federated learning to split learning. \\
         \hline
        ~\cite{qi2023model} & Aggregation in federated learning & Introduced model aggregation in federated learning, covering detailed taxonomy of aggregation methods. Also explored hot issues in federated learning. \\
         \hline
        ~\cite{sah2022aggregation} & Aggregation in federated learning & Discussed current aggregation techniques and challenges in federated learning. \\
         \hline
        ~\cite{moshawrab2023reviewing} & Aggregation in federated learning & Introduced several federated learning aggregation strategies and algorithms, also including limitations and future perspectives.\\
         \hline
    \end{tabular}
    \vspace{10pt}
    \label{tab_survey}
\end{table*}

The integration of distributed learning paradigms with wireless communication systems has gained significant attention in recent years. Several surveys have explored different aspects of distributed AI in wireless networks, focusing on frameworks, algorithms, privacy preservation, and resource allocation strategies. The topics and key contributions of related works are summarized in Table~\ref{tab_survey}. While these works provide valuable insights, they primarily focus on FL, edge AI, or aggregation mechanisms, with limited discussions on SL and AL for foundation models over 6G networks. In this section, we summarize the most relevant works and highlight their limitations, underscoring the necessity of this survey.

Several studies have investigated distributed learning techniques for wireless communication systems. Qian \textit{et al.}~\cite{qian2022distributed} introduced typical frameworks and algorithms for distributed learning, demonstrating their applicability across different network layers. Similarly, Chen \textit{et al.}~\cite{chen2021distributed} provided an overview of emerging distributed learning paradigms, including FL, and discussed their relevance to modern wireless networks. Extending this line of research, Hu \textit{et al.}~\cite{hu2021distributed} not only reviewed distributed learning techniques but also examined potential adversarial attacks and corresponding countermeasures, emphasizing security challenges in wireless AI deployment. These works establish a strong foundation for understanding distributed learning in wireless networks but do not specifically address the challenges of deploying foundation models with SL and AL.

In parallel, edge AI has emerged as a crucial paradigm for intelligent and adaptive 6G networks. Letaief \textit{et al.}~\cite{letaief2021edge} introduced an edge AI framework for 6G, discussing new design principles, resource allocation strategies, and system architectures. Shi \textit{et al.}~\cite{shi2020communication} further examined communication-efficient edge AI techniques, focusing on training and inference optimizations at the network edge. While these studies explore AI acceleration at the edge, they lack discussions on how SL can enhance AI model training efficiency by offloading computational workloads across edge and cloud layers.

With the increasing relevance of foundation models in AI-driven wireless networks, recent studies have explored their potential integration into 6G systems. Chen \textit{et al.}~\cite{chen2024big} provided an overview of big AI models for 6G, discussing opportunities, challenges, and future research directions. However, this work primarily focuses on model development and deployment rather than the specific training methodologies required for foundation models in distributed environments. The scalability, privacy, and computational constraints of training foundation models over wireless networks remain open challenges that our survey aims to address.

To bridge the gap between FL and SL, some studies have analyzed hybrid learning approaches. Duan \textit{et al.}~\cite{duan2022combined} explored the combination of FL and SL in edge computing, identifying open problems and future research directions. Similarly, Thapa \textit{et al.}~\cite{thapa2021advancements} investigated advancements in FL with a focus on privacy preservation, examining the transition from FL to SL. Although these studies provide insight into the combination of FL and SL, they do not explore AL or its impact on foundation models in wireless AI.

Another critical aspect of distributed AI is model aggregation in FL, which plays a fundamental role in optimizing model convergence and communication efficiency. Qi \textit{et al.}~\cite{qi2023model} introduced a taxonomy of aggregation methods in FL, identifying key challenges in decentralized learning. Sah \textit{et al.}~\cite{sah2022aggregation} and Moshawrab \textit{et al.}~\cite{moshawrab2023reviewing} further discussed aggregation techniques, challenges, and future perspectives, highlighting the need for scalable and efficient aggregation mechanisms in distributed AI. However, these works primarily focus on aggregation in FL, without considering aggregation in SL-based architectures, which is crucial for handling heterogeneous wireless environments and resource-constrained edge nodes.

Despite these valuable contributions, existing surveys lack a comprehensive analysis of SL and AL for foundation models over wireless communication networks. Furthermore, prior works have not explored how SL and AL can be integrated into AI-native 6G networks to optimize training efficiency, privacy preservation, and communication overhead. This survey aims to fill these gaps by systematically reviewing SL and AL techniques, their architectural designs, and their applicability in next-generation wireless networks. By bridging the intersection of distributed learning, foundation models, and 6G communication systems, our work provides a new perspective on AI-driven wireless intelligence.

\subsection{Motivation, Scope and Outline of the Paper}

Despite the growing interest in AI-driven wireless networks, existing surveys primarily focus on FL, with limited discussions on SL and AL. Moreover, there is currently no comprehensive taxonomy of SL and AL frameworks for large-scale AI model training, particularly in 6G wireless communication systems. Additionally, the interplay between wireless communication constraints and AI model optimization remains underexplored. Specifically, we explore the architectural designs and methodologies for SL and AL, analyze the role of wireless communication technologies in enabling distributed AI, and discuss key challenges, security concerns, and open research directions.

This survey provides a comprehensive analysis of SL and AL for Foundation Models in wireless communication systems, particularly in 6G networks. The key contributions of this work are summarized as follows:
\begin{itemize}
    \item We provide a comprehensive analysis of SL and AL frameworks, introducing various configurations like vanilla SL, extended SL, U-shape SL, and SL for vertically partitioned data, while exploring AL methodologies and aggregation techniques. This analysis covers their advantages, limitations, and applications in distributed AI training, with particular focus on their role in decentralized machine learning and scalable AI model training for wireless systems.
    \item  We examine how SL and AL can optimize the training and inference of large-scale foundation models in wireless environments, reducing computation and communication overhead. We analyze how pre-training, fine-tuning, and prompt-based adaptation enable scalable model deployment in 6G networks.
    \item We investigate the interplay between SL, AL, and wireless communication technologies, identifying communication bottlenecks and proposing methods to optimize data transmission efficiency. We discuss techniques such as activation compression, gradient sparsification, and asynchronous aggregation, reducing communication latency and energy consumption in large-scale wireless AI deployments.
    \item We investigate how SL and AL can be integrated with next-generation wireless technologies, such as joint communication-sensing-computation (JCSC), space-air-ground integrated networks (SAGINs), Semantic Communication, RIS, and Quantum Communication.
\end{itemize}

\begin{figure*}[t]
    \centering % 表示居中
    \includegraphics[width=0.73\linewidth]{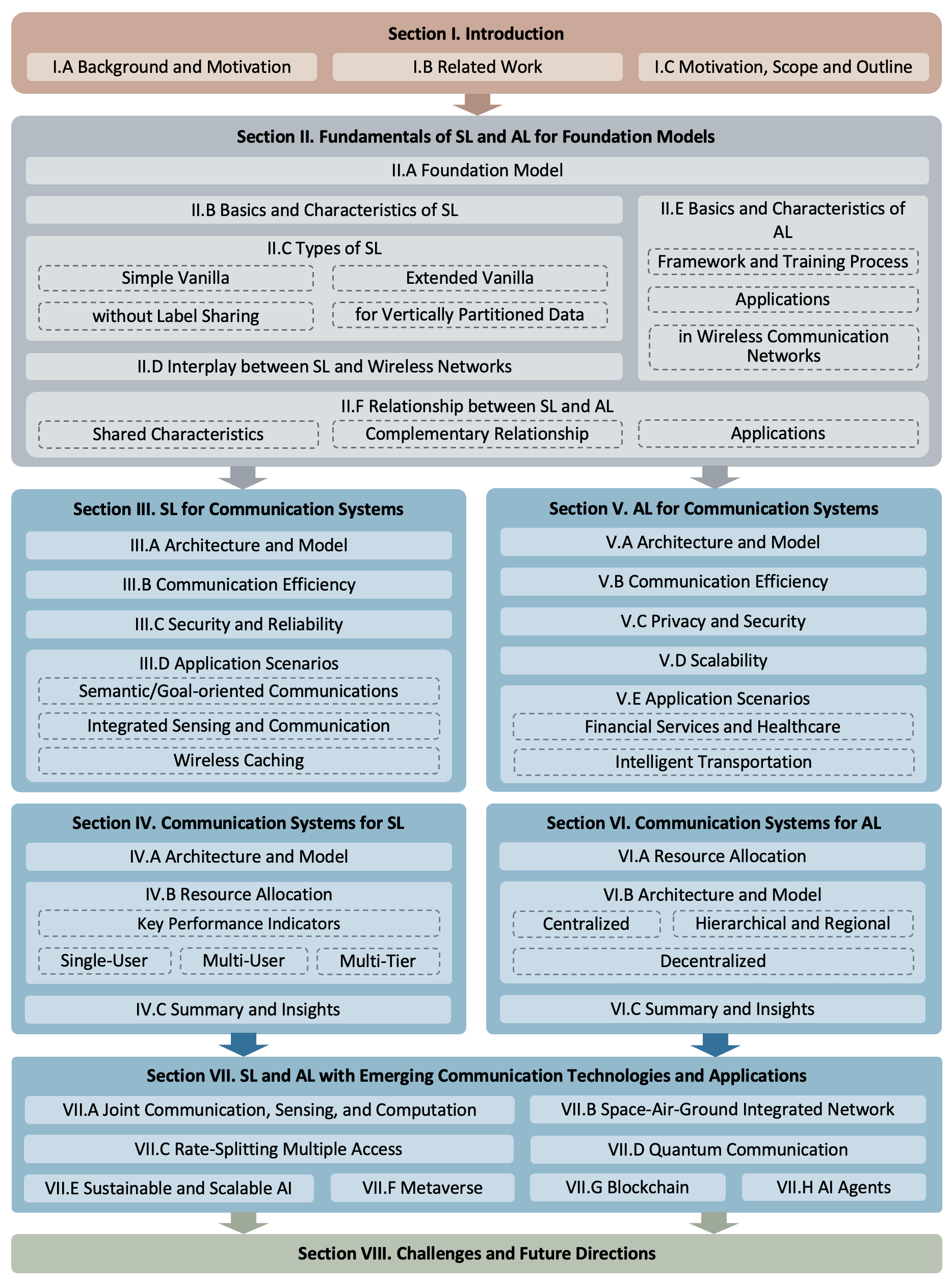}
    \caption{The outline of this survey.} 
    \label{fig_outline}
\end{figure*}

As the outline illustrated in Fig.~\ref{fig_outline}, the survey is organized as follows. Section~\ref{sec_fundamentals} examines the fundamentals of SL and AL for foundation models. Section~\ref{sec_SL} presents the SL for communication systems, and the communication systems are discussed in Section~\ref{sec_for_SL}. The AL for communication systems is shown in Section~\ref{sec_AL}. Section~\ref{sec_for_AL} addresses the communication systems for AL. Section~\ref{sec_emerging} explores emerging communication technologies and applications of SL and AL. Section~\ref{sec_conclusions} provides the conclusions.

\begin{table}[]
    \centering
    \renewcommand{\arraystretch}{1.5} % 设置行高为默认的1.5倍
    \caption{List of Common Abbreviations.}
    \begin{tabular}{|c|l|}
         \hline
         \textbf{Abbreviation} & \textbf{Description} \\
         \hline
         AL & Aggregation Learning \\
         \hline
         CAV & Connected and Autonomous Vehicle \\
         \hline
         DNN & Deep Neural Network \\
         \hline
         FL & Federated Learning \\
         \hline
         GAN & Generative Adversarial Network \\
         \hline
         IoT & Internat of Things \\
         \hline
         ISAC & Integrated Sensing and Communication \\
         \hline
         JCSC & Joint Communication-Sensing-Computation \\
         \hline
         MEC & Mobile Edge Computing \\
         \hline
         NLP & Natural Language Processing \\
         \hline
         RIS & Reconfigurable Intelligent Surfaces \\
         \hline
         SAGIN & Space-Air-Ground Integrated Network \\
         \hline
         SL & Split Learning \\
         \hline
         UAV & Unmanned Aerial Vehicle \\
         \hline
         URLLC & Ultra-Reliable Low-Latency Communications \\
         \hline
    \end{tabular}
    \vspace{10pt}
    \label{tab_abbr}
\end{table}

\section{Fundamentals of Split Learning and Aggregation Learning for Foundation Models}
\label{sec_fundamentals}

\begin{figure*}
    \centering % 表示居中
    \includegraphics[width=0.9\linewidth]{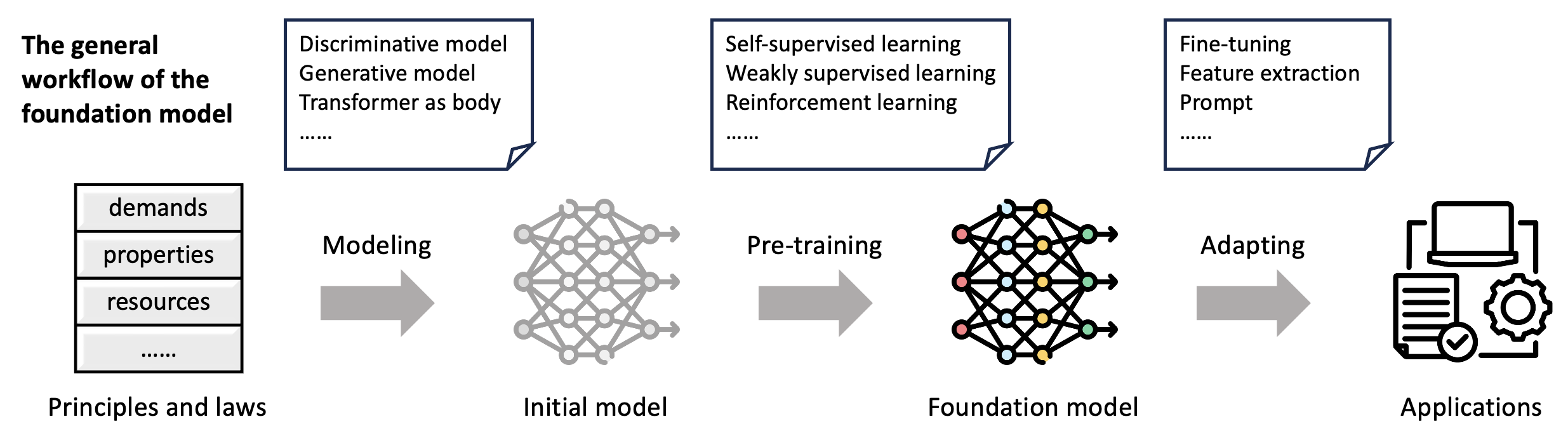}
    \caption{The general workflow of the foundation model. It captures broad-adaptable intelligence and provides numerous downstream applications through appropriate modeling, sufficient pre-training, and customized adaptations.} 
    \label{fig_large_ai}
\end{figure*}

\subsection{Fundamentals of Foundation Model}

Foundation models represent a transformative paradigm in machine learning, significantly advancing fields such as natural language processing (NLP), computer vision, and autonomous systems~\cite{yuan2021florence}. These models are characterized by large-scale architectures, extensive parameterization, and the ability to generalize across diverse tasks~\cite{bommasani2021opportunities}. The general workflow of a foundation model is illustrated in Fig.~\ref{fig_large_ai}. Unlike traditional deep neural networks (DNNs), which require task-specific training and lack generalizability, foundation models are pre-trained on massive datasets and subsequently adapted to specific applications through fine-tuning, few-shot, or even zero-shot learning. This capability enables foundation models to surpass the limitations of traditional AI by offering broad-adaptive intelligence rather than isolated, task-specific solutions~\cite{chen2024big}.

In 6G communication systems, foundation models have enormous potential to improve network intelligence, optimize resource allocation, and improve real-time decision-making. Traditional AI-driven wireless networks rely on manually engineered heuristic algorithms or task-specific deep learning models, which often fail to generalize across different network conditions, user demands, or spectrum environments. Using foundation models, 6G networks can achieve self-evolving intelligence, where a single pre-trained model can dynamically adapt to diverse network scenarios, including intelligent radio resource management, URLLC, and semantic communication~\cite{10333452,zhou2024comprehensive,10960269,10615635}.

Pre-training is a crucial feature of foundation models, distinguishing them from traditional machine learning paradigms. Instead of training AI models in isolation for each specific task, a foundation model is first pre-trained on large-scale datasets, often through collaboration between cloud and edge infrastructures. This pre-training enables the model to capture hierarchical representations, learn universal patterns, and develop robust generalization capabilities before being deployed in real-world applications. For 6G wireless communication systems, pre-training of foundation models offers several key advantages:
\begin{itemize}
    \item Improved Wireless Intelligence and Performance: By leveraging heterogeneous wireless datasets collected from IoT devices, mobile users, and edge nodes, foundation models can achieve superior intelligence for network optimization, traffic prediction, and intelligent spectrum sharing. These models dynamically adapt to varying conditions such as interference, mobility patterns, and congestion levels, thereby enhancing network efficiency.
    \item Reduced Overhead for Real-Time Adaptation: One of the major challenges in the deployment of wireless AI is the high cost of collecting, labeling, and training models for each new scenario. With foundation models, only a lightweight fine-tuning process or prompt-based adaptation is required to customize the model for different 6G applications. This drastically reduces the computational burden on resource-constrained devices while ensuring optimal performance across various network environments~\cite{chen2024big}.
\end{itemize}

Specifically, the adoption of foundation models in 6G networks enables a paradigm shift in wireless intelligence \cite{10579546}. Unlike current 5G AI solutions that primarily rely on shallow network automation, 6G networks will be AI-native, meaning that AI-driven optimization will be embedded into every layer of the wireless stack~\cite{jiang2021road,saad2019vision}. The key transformations facilitated by foundation models include:
\begin{itemize}
    \item  Integrated Functionality: Instead of developing separate AI models for channel estimation, beamforming, mobility prediction, and resource allocation, foundation models unify these functions within a single adaptable model, significantly enhancing efficiency and interoperability.
    \item Flexible Network Architectures: Traditional wireless systems often rely on rigid, centralized architectures, making them inefficient in handling dynamic environments. Foundation models enable collaborative and decentralized network management driven by AI, where base stations, edge nodes, and user devices work together to optimize network performance in real-time.
    \item Differentiated and Personalized Services: Future 6G networks will provide context-sensitive and user-specific services using foundation models to understand user behavior, application requirements, and environmental conditions. For example, an autonomous vehicle in a high-density urban area will require ultra low-latency connections, while a remote industrial sensor may prioritize energy efficiency over speed. Foundation models allow for adaptive service differentiation, ensuring optimal performance tailored to individual user needs.
\end{itemize}

Foundation models are poised to redefine AI-driven wireless networks, offering generalizable, efficient, and scalable intelligence for 6G communication systems. By leveraging pre-training, fine-tuning, and prompt engineering, these models will enable flexible, adaptive, and highly optimized network management. As 6G networks move toward full AI-native architectures, the integration of foundation models will be a key enabler of next-generation wireless intelligence~\cite{tataria20216g,akyildiz20206g}.

\begin{figure*}
    \centering % 表示居中
    \includegraphics[width=1\linewidth]{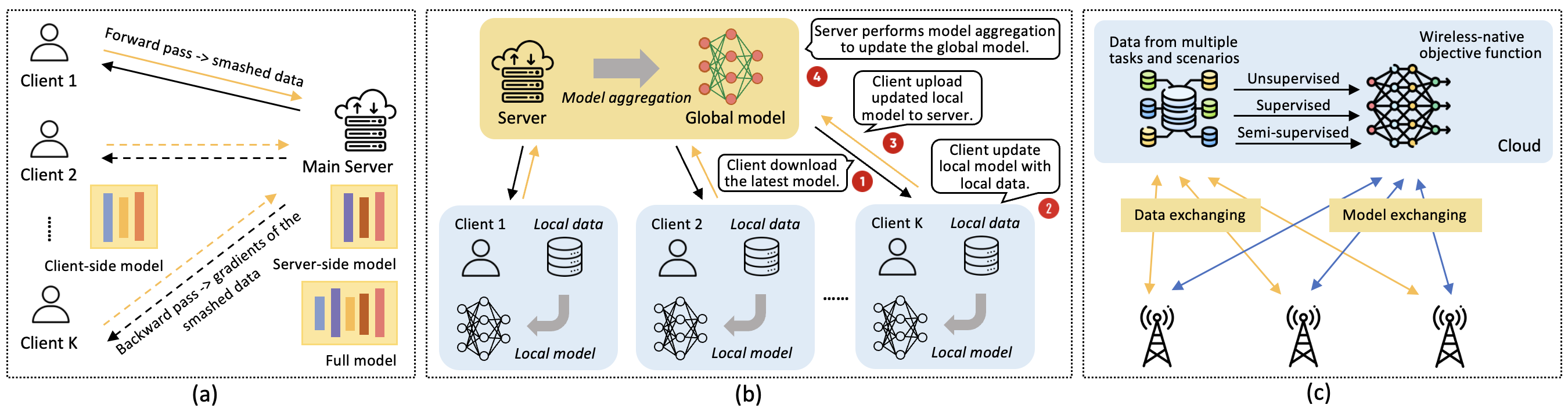}
    \caption{(a) A framework architecture of multi-agent SL. (b) A framework of model aggregation. (c) An ML framework for wireless communications.} 
    \label{fig_framework}
\end{figure*}

\subsection{Basics and Characteristics of Split Learning}

SL partitions a neural network between multiple entities, typically clients (e.g., user devices) and servers (e.g., base stations or cloud data centers). The client-side model ($W_C$) runs on a user device, handling initial computations, while the server-side model ($W_S$) processes deeper layers with higher computational demands. The overall framework of SL is illustrated in Fig.~\ref{fig_SL_setup}. The training process consists of two key steps:

\begin{figure}
    \centering % 表示居中
    \includegraphics[width=1\linewidth]{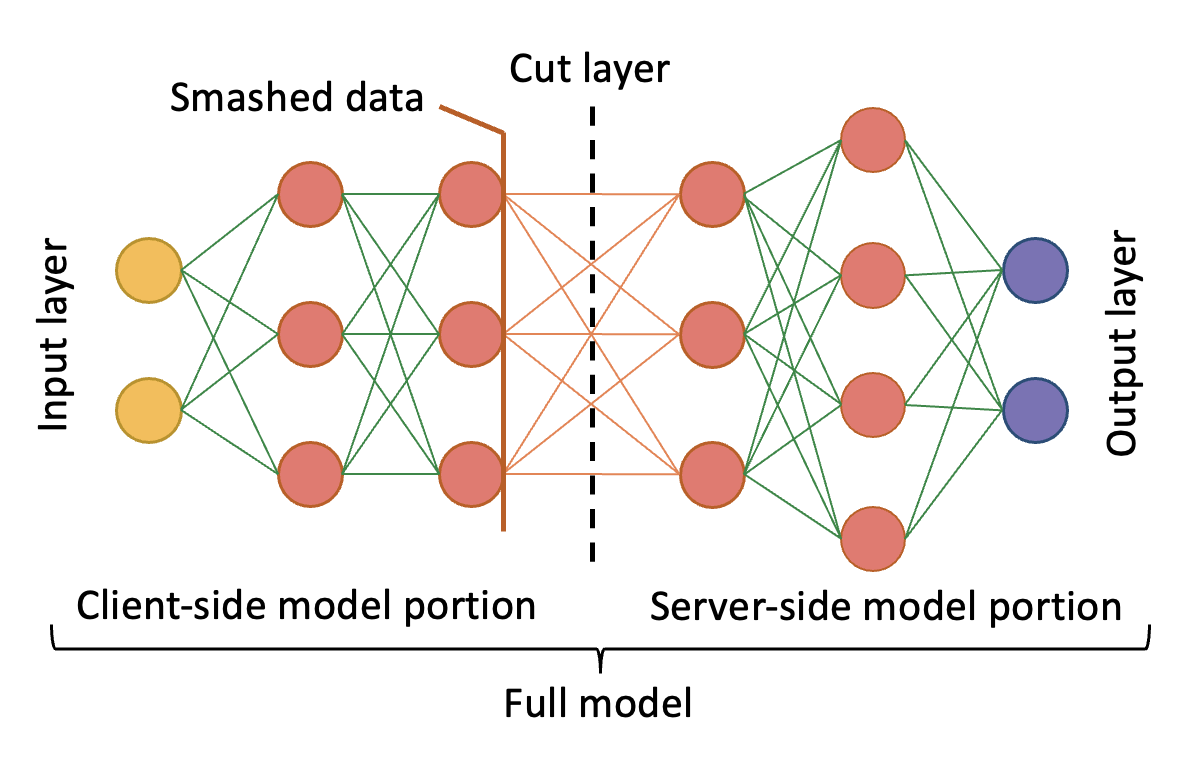}
    \caption{A simple setup of SL, where the neural network is divided into two parts: the client-side model $W_C$ and the server-side model $W_S$.} 
    \label{fig_SL_setup}
\end{figure}

\begin{itemize}
\item \textbf{Forward Propagation:} The client processes its local data through $W_C$ up to the \textit{cut layer}, producing smashed data (intermediate activations). This smashed data, often accompanied by labels, is then transmitted to the server. The server takes this as input, completing forward propagation through $W_S$.
\item \textbf{Backward Propagation:} The server computes the loss and performs backpropagation, updating $W_S$. The gradients of the smashed data are sent back to the client, where they are used to update $W_C$. This process continues iteratively until the model converges.
\end{itemize}

By splitting the model in this manner, SL enables efficient training on edge devices while leveraging the computational power of centralized servers.

In SL frameworks involving multiple clients, such as the setup shown in Fig.~\ref{fig_framework}(a), clients train their portion of the model in a sequential round-robin fashion, alternating between training epochs~\cite{thapa2021advancements}. To ensure a globally consistent model, synchronization mechanisms are required:
\begin{itemize}
    \item Centralized Synchronization: Clients upload their trained model weights to a central server, which then distributes them to other clients.
    \item Peer-to-Peer Synchronization: Clients communicate directly, downloading the latest model updates from the previous participant~\cite{gupta2018distributed}.
\end{itemize}
Without successful synchronization, SL frameworks may experience inconsistent training progress, potentially leading to unstable or non-converging models~\cite{thapa2021advancements}.

In 6G wireless networks, SL demonstrates potential applications, where distributed AI and edge computing are key components. Research has identified three potential benefits: First, it enables resource allocation optimization by distributing computational tasks between edge devices and servers. Second, it provides a degree of privacy protection by transmitting transformed activations instead of raw data. Third, it offers the potential for scaling across distributed networks, which could support various IoT applications. These characteristics suggest SL could be valuable for model training in resource-constrained IoT environments, though further research is needed to fully validate its effectiveness in real-world deployments.

\subsection{Types of Split Learning}

\begin{figure*}[t]
    \centering % 表示居中
    \includegraphics[width=1\linewidth]{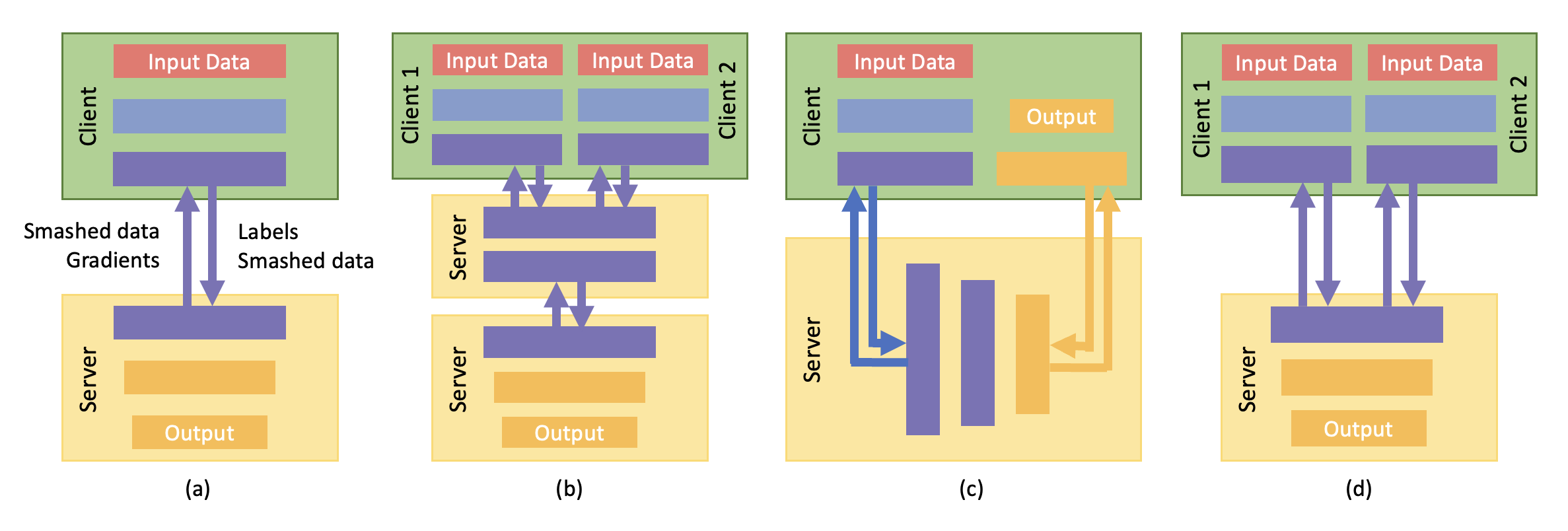}
    \caption{Configurations of SL (a) simple vanilla, (b) extended vanilla, (c) without label sharing, and (d) vertically partitioned data.}
    %, (e) multi-task output, and (f) multi-hop.} 
    \label{fig_SL_configuration}
\end{figure*}

There are several configurations for an SL framework applied to foundation models, as depicted in Fig.~\ref{fig_SL_configuration}, namely \textit{simple vanilla split learning}, \textit{extended vanilla split learning}, \textit{multi-hop split learning}, and \textit{split learning without label sharing}~\cite{vepakomma2018split}. Their summary and comparison are shown in Table~\ref{tab_SL_config}.
%, \textit{split learning for vertically partitioned data}, and \textit{split learning for multi-task output}

\begin{table*}[t]
    \centering
    \renewcommand{\arraystretch}{1.5} % Adjust row height for readability
    \caption{Summary of SL Configurations.}
    \begin{tabular}{|m{8cm}|m{8cm}|}
         \hline
         \textbf{Simple Vanilla SL} & \textbf{Extended Vanilla SL} \\
         \hline
         
         \textbf{Description:} The most straightforward configuration. The neural network model is split into two portions for a pair of client and server that share the smashed data and labels. 
        
         \textbf{Pros:} 
         \begin{itemize}
             \item Easy to implement and manage as the basic form of SL~\cite{chen2021communication}.
         \end{itemize} 
        
         \textbf{Cons:} 
         \begin{itemize}
             \item Labels are sent to the server, which could compromise privacy.
             \item Frequent communication between client and server, increasing latency.
         \end{itemize} 
        
         & 
        
         \textbf{Description:} The model is split into multiple portions, where intermediate layers are processed by other workers before passing to the main server. 
        
         \textbf{Pros:}  
         \begin{itemize}
             \item Reduces back-and-forth communication overhead.
             \item Enhances privacy protection by limiting direct label exposure~\cite{nam2023active}.
             \item More scalable for larger datasets and multiple clients.
         \end{itemize}
        
         \textbf{Cons:}  
         \begin{itemize}
             \item Still shares labels with the server, leading to potential privacy risks.
         \end{itemize} \\
         
         \hline
         
         \textbf{SL without Label Sharing} & \textbf{SL for Vertically Partitioned Data} \\
         \hline
         
         \textbf{Description:} The client only transmits smashed data to the server without sharing corresponding labels. 
        
         \textbf{Pros:}  
         \begin{itemize}
             \item Both data and labels remain private, enhancing security~\cite{gupta2018distributed}.
         \end{itemize}
        
         \textbf{Cons:}  
         \begin{itemize}
             \item Clients must compute loss and backpropagate gradients locally.
             \item Increases computational burden on resource-constrained devices.
         \end{itemize} 
        
         & 
        
         \textbf{Description:} Each client owns a subset of features of the sample data. Each client performs forward propagation locally and shares only intermediate activations with the server.  
        
         \textbf{Pros:}  
         \begin{itemize}
             \item No single client or server has full data access, enhancing privacy~\cite{hardy2017private, navathe1984vertical}.
             \item Useful when multiple parties have complementary datasets.
         \end{itemize} 
        
         \textbf{Cons:}  
         \begin{itemize}
             \item Requires precise synchronization among clients.
             \item Complexity in data management and model coordination.
             \item If one client drops out, training efficiency is severely affected.
         \end{itemize} \\
         
         \hline
    \end{tabular}
    \vspace{10pt}
    \label{tab_SL_config}
\end{table*}

\subsubsection{Simple Vanilla SL}

Simple vanilla SL is the most fundamental configuration, as shown in Fig.~\ref{fig_SL_configuration}(a). In this setup, a neural network is split into two parts: a client-side model and a server-side model. The client processes data up to a specific cut layer, producing \textit{smashed data}, which is then transmitted to the server along with the corresponding labels. The server completes forward propagation, computes the loss, and backpropagates the gradients to the client. Several studies have leveraged this basic SL framework to enhance privacy and efficiency in distributed AI. In~\cite{chen2021communication, ayad2021improving}, threshold mechanisms were introduced to train deep neural networks collaboratively between servers and clients without sharing raw data, thereby ensuring privacy and security in wireless communication systems.

\subsubsection{Extended Vanilla SL}

Extended vanilla SL builds upon the basic framework by introducing additional processing layers between clients and the main server, as depicted in Fig.~\ref{fig_SL_configuration}(b). These intermediary processing nodes help optimize communication efficiency and improve privacy protection. For instance,~\cite{nam2023active} introduces a \textit{tiny server}, which selectively transmits smashed data to the main server based on its informativeness. This approach reduces communication overhead by avoiding unnecessary data exchanges. Similarly,~\cite{oh2024privacy} proposes a \textit{mixer}, an entity that injects Gaussian noise into smashed data and randomly mixes patches across clients. This privacy-enhancing mechanism helps mitigate membership inference attacks, making SL more secure, especially for large-scale vision transformer models.

\subsubsection{SL without Label Sharing (U-shape SL)}

SL without label sharing, also known as U-shape SL, is depicted in Fig.~\ref{fig_SL_configuration}(c). Unlike vanilla SL, where labels are shared with the server, this configuration ensures that labels remain on the client side, enhancing privacy. The workflow is modified as follows:
\begin{itemize}
    \item The client performs forward propagation and transmits smashed data to the server.
    \item The server processes its portion of the network up to a designated \textit{server cut layer} and sends the activations back to the client.
    \item The client completes forward propagation up to the final output layer, computes the loss, and starts backpropagation.
    \item Gradients are transmitted back to the server for further training, forming a U-shape data flow.
\end{itemize}
This method is particularly useful in privacy-sensitive domains such as healthcare, where labels (e.g., disease status of patients) contain highly confidential information. As demonstrated in~\cite{gupta2018distributed}, U-shape SL significantly reduces the risk of data leakage while maintaining efficient model training.

\subsubsection{SL for Vertically Partitioned Data}

This configuration, illustrated in Fig.~\ref{fig_SL_configuration}(d), enables multiple institutions holding different modalities of sensitive data to train a shared model without exchanging raw data~\cite{hardy2017private, navathe1984vertical}. Unlike previous configurations, each client processes only a subset of features rather than complete input samples. The process follows these steps:
\begin{itemize}
    \item Each client carries out forward propagation on its local model using its specific data modality.
    \item The clients send their smashed data to the server, which concatenates all received activations and continues forward propagation.
    \item During backpropagation, the server computes the gradients, splits them accordingly, and sends them back to the respective clients.
    \item Each client updates its local model independently.
\end{itemize}

This setup is ideal for privacy-preserving collaborative learning, where multiple entities contribute complementary datasets (e.g., hospitals sharing medical imaging and patient records) while maintaining strict confidentiality.

\subsubsection{Summary of SL Configurations}

The diverse configurations of SL offer scalability and flexibility to meet the varying needs of distributed AI systems. Each approach is designed to address specific challenges:
\begin{itemize}
    \item Simple Vanilla SL: Basic and easy to implement but requires label sharing.
    \item Extended Vanilla SL: Improves efficiency and privacy through intermediate processing layers.
    \item SL without Label Sharing (U-shape SL): Eliminates label exposure, making it suitable for highly sensitive applications.
    \item SL for Vertically Partitioned Data: Allows collaborative learning across institutions without direct data sharing.
\end{itemize}

These configurations demonstrate the adaptability of SL in achieving privacy preservation, computational efficiency, and large-scale AI deployment. While these represent some of the most common SL frameworks, other variations exist, making SL a highly customizable approach for future AI-driven wireless networks.

\subsection{The Interplay between SL and Wireless Networks}

The rise of foundation models has driven the need for distributed training and real-time inference, shaping next-generation wireless networks. As we move toward 6G, the focus shifts from \textit{``connected things"} to \textit{``connected intelligence"}~\cite{tong20226g}. However, deep learning models demand high computation and communication resources, leading to challenges such as latency, energy consumption, network congestion, and privacy risks. 

6G networks introduce several innovations that enhance SL~\cite{letaief2021edge}. URLLC enables real-time model training and inference~\cite{feng2021ultra}. AI-native networks dynamically allocate bandwidth and computing resources for SL workloads. RIS improves wireless channel efficiency, ensuring reliable smashed data transmission~\cite{liu2021reconfigurable}. ISAC helps optimize SL training by collecting real-time network data~\cite{liu2022integrated}. Edge Intelligence further supports decentralized SL, reducing reliance on cloud computing~\cite{shi2020communication}. SL plays a key role in 6G edge AI, enabling collaborative training across edge devices and servers. SL frameworks—such as vanilla, extended, and vertically partitioned SL—can be tailored to different privacy and efficiency requirements. By processing data locally and sharing only activations, SL enhances privacy, security, and scalability while minimizing network congestion.  Beyond training, SL also improves edge inference by enabling real-time, low-latency AI decision-making. It reduces the computational burden on resource-constrained devices, ensuring fast, efficient, and adaptive AI services. In autonomous systems, smart cities, and industrial automation, SL combined with 6G ensures intelligent, privacy-preserving, and real-time AI solutions.

\subsection{Basics and Characteristics of Aggregation Learning}

In machine learning, AL is a broad concept that generally refers to any process of information fusion. In the context of distributed learning discussed in this paper, AL typically refers to the process of aggregating model updates from multiple clients to optimize the performance of the global model, while simultaneously reducing communication overhead and improving computational efficiency. In certain application scenarios, it can also enhance data privacy protection~\cite{laskaridis2020spinn}. With the rise of 5G, 6G, IoT, and mobile edge computing (MEC), efficient learning methods are needed to address communication constraints, privacy risks, and computational limitations~\cite{bonawitz2019towards,yang2019federated}. Traditional centralized approaches make it difficult to meet the demands of efficient learning in distributed wireless environments due to high communication overhead and significant data privacy risks~\cite{niyato2017survey}. AL provides a scalable solution by allowing devices to train local models and asynchronously aggregate their updates at a central server, ensuring low-latency, privacy-preserving, and energy-efficient AI training~\cite{zhao2018federated}.

\subsubsection{The Framework and Training Process of AL}
In the AL framework, multiple clients (such as devices, sensors, or nodes) independently train local models and send local model updates (such as gradients or weights) to a central server, as shown in Fig.~\ref{fig_AL_structure}. The server aggregates updates from each client to form a global model and returns it to the clients. Unlike centralized learning, AL emphasizes sharing model updates, rather than raw data, between different devices or nodes, thereby protecting data privacy~\cite{qi2023model}. The common training process is as follows:

\begin{figure}
    \centering
    \includegraphics[width=1\linewidth]{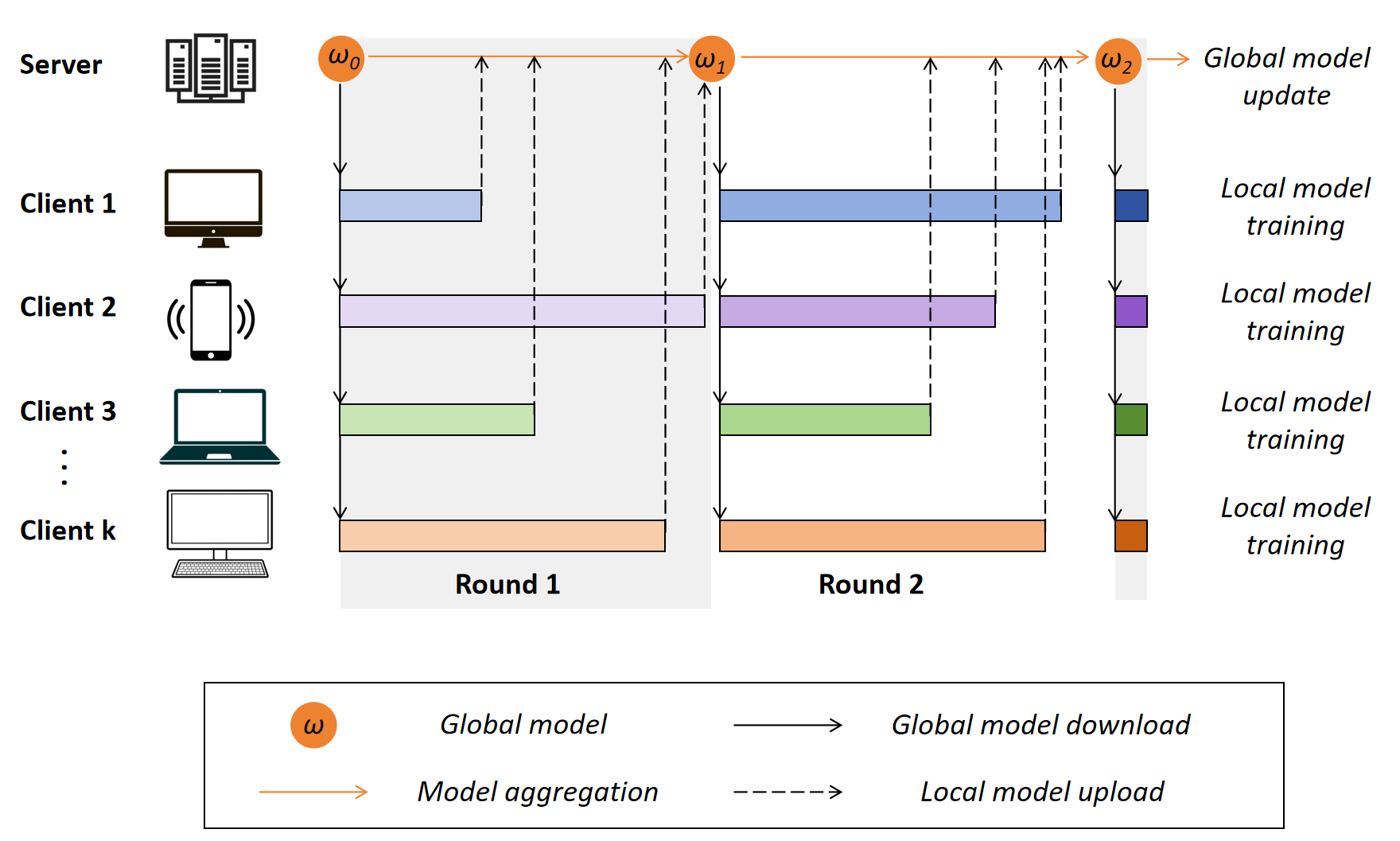}
    \caption{\fontsize{9.5pt}{10pt}\selectfont The common framework of AL. The server distributes the global model at the start of each round, and clients asynchronously train and upload updates, which are partially aggregated to form the new global model.}
    \label{fig_AL_structure}
\end{figure}

\begin{itemize}
    \item Local Training: Each client trains its local model on its private dataset, generating local model updates. These updates are obtained through some form of computation (such as gradient descent).

    \item Model Aggregation: Each client sends its local updates to the central server. The server aggregates these updates according to certain rules (such as weighted averaging or summation) to form a global model.

    \item Update and Iteration: The aggregated global model is returned to the clients, and they perform local training again. This process continues iteratively until the global model converges.
\end{itemize}

\subsubsection{The Applications of AL}
In distributed learning, AL is often used in conjunction with other techniques, such as ensemble learning and federated learning, to optimize the model training process.

AL is commonly integrated with ensemble learning, particularly when multiple model outputs need to be aggregated. In ensemble learning, multiple models are trained independently and generate predictions, which are then combined using methods such as voting or weighted averaging to improve prediction accuracy and model robustness~\cite{zhou2012ensemble}. AL plays a crucial role in this process by aggregating updates from different models, and integrating local updates from multiple models to form a global model. Specifically, AL applies weighted or other aggregation methods to optimize the global model, enabling it to leverage the strengths of different models and enhance its generalization ability on new data~\cite{lu2024merge}.

FL is a specific application of AL. FL is characterized by its emphasis on data privacy protection and its client-server architecture, allowing multiple clients to train models locally while sending only model updates (rather than raw data) to a central server for aggregation. AL plays a crucial role in FL by aggregating model updates from clients, optimizing the global model to enhance both privacy protection and model accuracy~\cite{qi2023model}.

Traditional FL relies on synchronous updates, making it susceptible to the straggler problem, where slower devices delay the global model update process~\cite{ruder2016overview,lee1999learning}. AL overcomes this limitation by enabling continuous model aggregation, making it more suitable for dynamic networks such as smart cities, autonomous driving systems, and real-time industrial IoT applications~\cite{chen2019machine}. Additionally, AL can incorporate encryption and secure multi-party computation techniques, ensuring that even if the central server receives updates from multiple devices, it can only access the aggregated results without viewing individual device updates.

\subsubsection{AL in Wireless Communication Networks}
AL provides several benefits for wireless communication networks. By transmitting model updates instead of raw data, it significantly reduces communication overhead, making it ideal for bandwidth-constrained environments~\cite{mcmahan2016federated}. Its scalability and flexibility make it suitable for large-scale, privacy-sensitive applications such as smart healthcare, industrial automation, and edge AI deployments~\cite{luong2019applications}. Furthermore, AL enhances privacy by keeping sensitive user data localized, reducing the risks associated with data breaches and unauthorized access~\cite{finn2017model}.

Despite its advantages, AL faces several challenges in practical applications. Model bias can affect fairness and generalization, particularly when client data is highly imbalanced. Communication overhead remains a concern, as large-scale participation can still lead to bandwidth congestion and latency~\cite{konecny2016federated}. Integrating AL with 6G requires optimizing aggregation efficiency and ensuring stability in dynamic environments~\cite{tataria20216g}. Furthermore, while asynchronous updates enhance adaptability, they may slow convergence. Continued research is essential to refine aggregation strategies and improve AL’s efficiency in 5G and 6G networks~\cite{wang2022asynchronous}.

In the era of 6G wireless networks, AL has emerged as a critical technology for enabling efficient distributed AI and edge computing. Unlike traditional approaches that transmit raw data, AL focuses on aggregating model updates, reducing bandwidth consumption and effectively alleviating communication bottlenecks in large-scale networks~\cite{tan2021integrated}. Additionally, AL enhances privacy protection by keeping sensitive information on local devices and only sharing encrypted model updates. Its asynchronous update capability improves system scalability, making AL particularly suitable for heterogeneous IoT environments with diverse computational and connectivity constraints~\cite{qi2023model}. As AL technology continues to evolve, further research is essential to validate its practical applicability and optimize its integration within 6G networks.

\subsection{The Relationship between SL and AL in Modern Communication Systems}

Although SL and AL differ in their implementation, their common goal is to enhance the training efficiency in large-scale distributed systems while ensuring effective data privacy protection. Therefore, exploring the relationship between SL and AL is of significant theoretical and practical value, as it helps in understanding their collaborative role in modern communication systems.

\subsubsection{Shared Characteristics}

SL and AL share common characteristics in several aspects, particularly in privacy protection, distributed computing, and communication efficiency optimization. 
\begin{itemize}
    \item SL and AL avoid the direct transmission of raw data, ensuring data privacy from different angles. In SL, the client only needs to transmit intermediate activation values rather than sensitive raw data; whereas in AL, devices only transmit updates of the local model, rather than raw data or original training samples~\cite{letaief2019roadmap,konecn2016federated}.
    \item Both optimize computational resources through distributed computing by allocating computational tasks to edge devices and servers, thereby alleviating the computational burden on individual devices, especially those with limited resources~\cite{konecn2016federated}.
    \item Both offer significant advantages in optimizing communication efficiency. By reducing the amount of data that needs to be transmitted, they effectively lower communication overhead, making them particularly suitable for high-performance wireless networks like 6G, where communication bandwidth is limited and low latency is crucial~\cite{navathe1984vertical}.

\end{itemize}

Therefore, SL and AL, while ensuring data privacy, improve the efficiency of distributed AI training through the optimization of both computational and communication resources.

\subsubsection{The Complementary Relationship between SL and AL}

SL and AL complement each other in several aspects, allowing for the combination of their strengths to further enhance the efficiency and security of distributed AI training. In terms of privacy protection, SL provides stronger data privacy since it only transmits intermediate activation values instead of model parameters or raw data. On the other hand, AL improves computational efficiency by asynchronously aggregating model updates from multiple devices, reducing training latency and communication overhead~\cite{letaief2019roadmap}. Therefore, SL and AL can complement each other at different levels, forming an efficient and secure distributed learning framework.

Specifically, SL can focus on data partitioning and feature extraction, assigning preliminary computation tasks to clients, thus reducing the amount of data to be transmitted and ensuring privacy. AL, on the other hand, can perform model updates and aggregation among multiple clients to optimize the global model’s performance, enhancing training efficiency~\cite{gupta2018distributed}. This combination can fully leverage the strengths of both SL and AL in various application scenarios, ensuring data privacy while minimizing computational and communication costs.

In future 6G networks, the synergy between SL and AL will realize even greater potential. SL can be applied to initial computations on edge devices, reducing reliance on cloud resources and safeguarding data privacy, while AL can aggregate training results from multiple edge devices during global model training to improve model generalization and optimize the training process. Through this collaborative approach, SL and AL will drive the efficient development of distributed AI in future wireless communication systems~\cite{letaief2021edge}.

\subsubsection{Applications in Modern Communication Systems}

SL and AL offer great potential in modern communication systems, especially in environments like 6G, IoT, and smart edge computing, where they can optimize computational resources and enhance data privacy. These technologies improve AI training efficiency and secure data handling in high-performance networks.

\begin{itemize}
    \item In smart healthcare, SL can process sensitive patient data locally at hospitals, transmitting only intermediate activations for further processing. Meanwhile, AL can aggregate models from multiple hospitals, improving generalization without compromising privacy~\cite{letaief2021edge}.
    \item In IoT applications, SL reduces data transmission by offloading computations to edge devices, while AL aggregates model updates from these devices to optimize training efficiency. This setup ensures data privacy while minimizing network congestion~\cite{mcmahan2017communication}.
    \item In 6G networks, the combination of SL and AL allows for real-time AI-driven applications like autonomous vehicles, smart cities, and industrial automation. SL handles local data processing, while AL aggregates updates to continuously refine the global model, making AI systems more efficient and scalable~\cite{letaief2019roadmap}.
\end{itemize}

Together, SL and AL enhance resource utilization and safeguard privacy, making them ideal for next-generation communication systems. They share similarities while also complementing each other, together enhancing the efficiency of AI training in modern communication systems. By combining the privacy protection and distributed computing strengths of both methods, they provide a more efficient and secure approach to AI model training. Future research can explore how to further optimize the integration of SL and AL, improving AI training performance in wireless communication environments, and enabling the development of more advanced, scalable AI systems in next-generation networks.

\section{Split Learning for Communication Systems} % Chen
\label{sec_SL}

In 6G wireless communication, SL offers significant benefits in terms of resource efficiency, privacy, and scalability, making it a promising solution for deploying foundation models across numerous mobile and IoT devices.

\subsection{Architecture and Model}

\begin{figure}[t]
    \centering % 表示居中
    \includegraphics[width=1\linewidth]{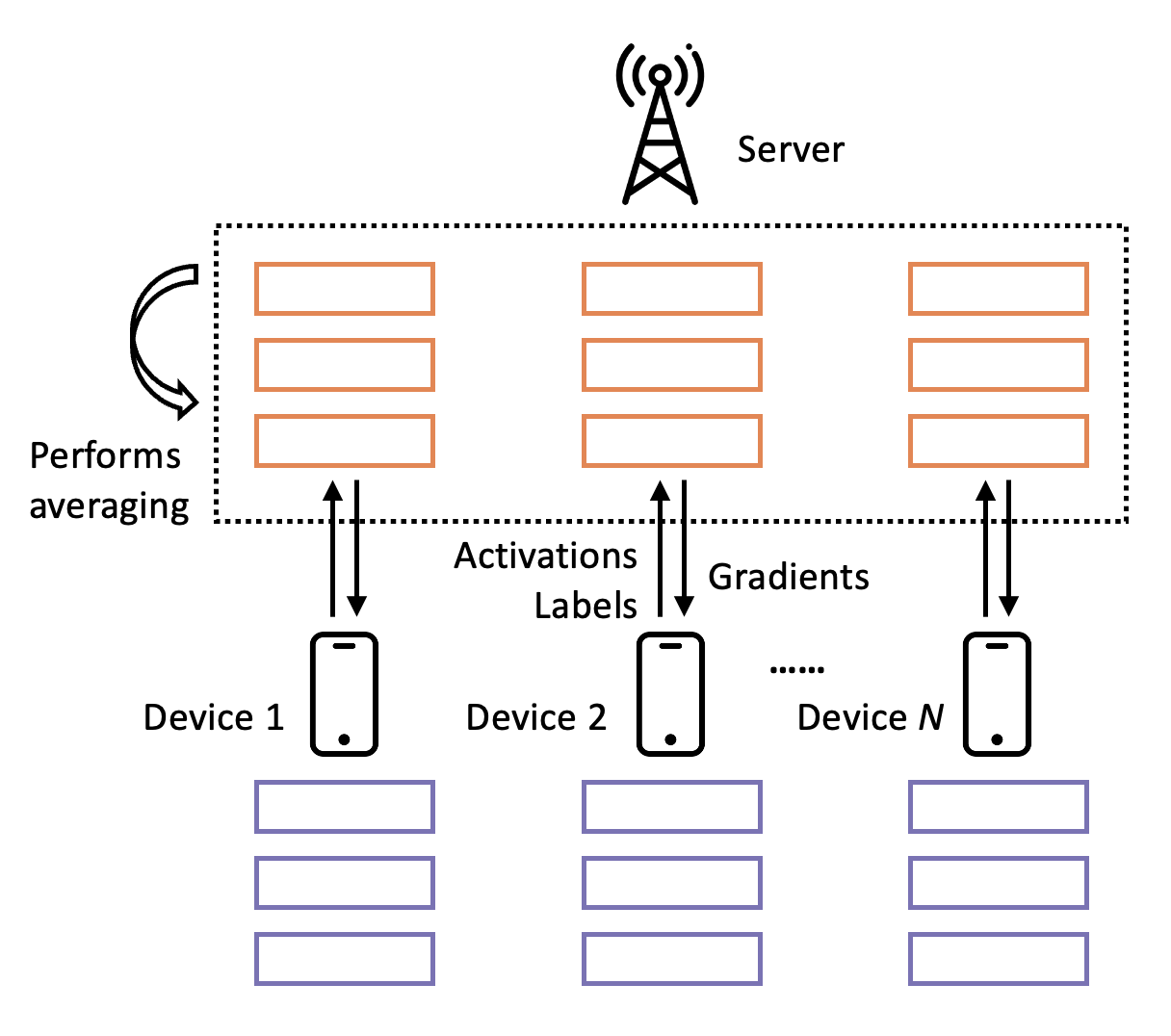}
    \caption{An illustration of the SL system over wireless networks.} \label{fig_SL_illustration}
\end{figure}

SL is an emergent distributed learning framework that can mitigate the overhead of foundation models' computation and wireless communication. In SL, the network model is split into two parts at the cut layer: a device-side model and a server-side model, as shown in Fig.~\ref{fig_SL_illustration}. The devices and the server communicate over a wireless channel. A device only needs to train its assigned model and transmit the intermediate representations (activations) of the cut layer to the server. The server, which has more computing resources, then utilizes the received information to train the remaining model~\cite{kim2023bargaining}.

\subsection{Communication Efficiency}

Communication efficiency is a critical concern due to the limited battery life of mobile and IoT devices in wireless communication systems \cite{11432118}. One main advantage of SL is that it can significantly reduce the computational load and energy consumption on client devices. In SL, only the initial layers of the foundation model network are processed on client devices. This reduces the amount of local computation, leading to lower energy consumption compared to traditional FL, where the entire model is trained locally. Besides, instead of transmitting raw data, clients only send activations to the server. These activations are often smaller than the raw data, reducing the communication overhead and energy usage associated with data transmission.

Although SL can overcome the disadvantage of computation-constrained on-edge devices, communication overhead is still a key bottleneck which in turn leads to difficulties in deployment on heterogeneous devices~\cite{chen2021communication,ayad2021improving,koda2020communication}. Previous studies ignored the fact that IoT networks or mobile edge networks consist of thousands of devices with bandwidth restrictions and time-varying communication channels~\cite{krouka2021communication}. The ideal communication system between edge devices and the centralized server is assumed to be functioning well. Additionally, the synchronization of local states across devices presents a challenge to SL networks. To enhance the communication efficiency of the SL system, it is necessary to considerably reduce the amount of activations and gradients transferred between edge devices and the server. In order to alleviate communication overhead in SL networks, two systematic approaches can be adopted: (i) reducing the number of forward and backward propagation rounds or (ii) reducing the size of smashed data in each round of communication.

\begin{figure*}
    \centering % 表示居中
    \includegraphics[width=1\linewidth]{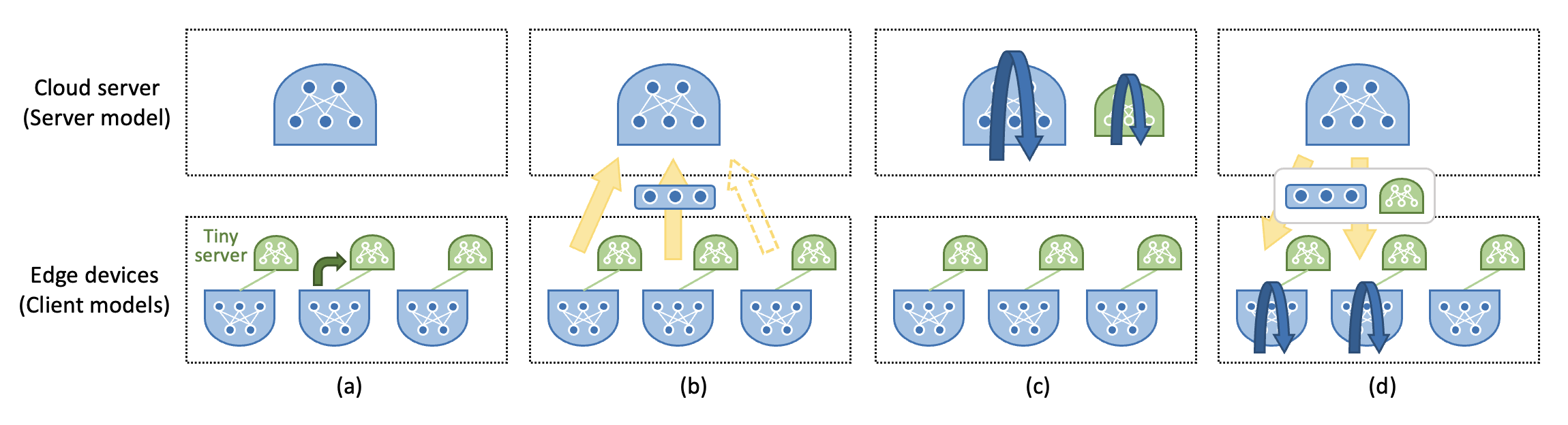}
    \caption{Training steps of SL with tiny server. (a) Forward propagation of clients and estimate loss via tiny server. (b) Devices send only informative smashed data with a estimated loss to the server. (c) Training the server and tiny server with received smashed data. (d) The server sends the gradient of cut layer and tiny server weights to each device.} 
    \label{fig_SL_tiny}
\end{figure*}

\textit{i) Reducing the frequency of data exchange:} In each round, agents send their local computation results to the centralized server to continue the forward propagation task and then gradients are transferred from the server to devices. As a result, the communication overhead increases when the number of devices grows~\cite{krouka2021communication}, which hinders tremendous scalability. Therefore, reducing the frequency of data exchange between the server and edge devices is a direct approach to prevent communication overhead. To mitigate the frequent communication cost, the authors in~\cite{nam2023active} introduce a novel SL framework with a tiny server as shown in Fig.~\ref{fig_SL_tiny}. The tiny server examines informative data at devices to selectively transmit smashed data to the server for training. This study utilizes a spatio-temporal distillation method, which allows the tiny server to evaluate similarly to the server's perspective. The proposed SL framework improves model performance while reducing overall communication costs by 50\%. In~\cite{chen2021communication,ayad2021improving}, the authors proposed a loss threshold that determines whether to exchange activation and gradient. At the end of the forward propagation, the centralized server quantifies the total loss and compares it with a predefined threshold. If the total loss is greater than the threshold, the gradient from the server is transferred to the clients to perform backward propagation on the rest of the network. In contrast, if the loss calculated at the server is lower than the threshold, there is no exchange of data between the server and clients. The server will keep track of gradients and carry out forward and backward cycles on their own network. Simulation results on VGG11, VGG13, and ResNet18 models on CIFAR-10 show that the communication cost is reduced by 1.64x-106.7x and the computations in the client are reduced by 2.86x-32.1x when the accuracy degradation is less than 0.5\% for the single-client case~\cite{chen2021communication}.

\textit{ii) Reducing the volume of data exchange:} Besides the mentioned approach, modern techniques such as Autoencoder~\cite{ayad2021improving}, Quantization, and Sparsify Activations and Gradients~\cite{chen2021communication} are utilized to reduce the volume of data transfer per wireless communication epoch. The authors in~\cite{ayad2021improving} investigated an SL network model equipped with an Autoencoder. Autoencoder is a special type of unsupervised artificial neural network that learns to compress the data while minimizing the reconstruction error. The encoder resides at the client side and is in charge of compressing the local gradient of the client before transferring to the server. At the server side, the decoder recovers the received gradient to the reconstructed version that will be performed forward process on the rest of the DNN. On the other hand,~\cite{chen2021communication} exploited quantizing the activation and gradient from 32-bit floating point to 8-bit floating point before transmission without much accuracy degradation. A search-based quantization scheme is implemented to search for the best combination of exponent bits and bias to quantize the activation and gradient before dispatch. The modern quantization approaches significantly cut back the data size in communication. However, quantization and sparsification methods typically result in massive compression errors because quantization is sensitive to outliers (top values in the feature map), and sparsification produces errors due to filtered values. Compression errors critically hinder SL performance. Compressing feature maps within SL leads to biased gradients that can negatively impact the convergence rates and diminish the generalization capabilities of the resulting large-scale models~\cite{beznosikov2023biased}. To address these challenges, the authors in~\cite{zhou2024mask} employ a narrow bit-width encoded mask to compensate for the sparsification error without increasing the order of time complexity. The theoretical and empirical evidence shows that the proposed mask-encoded sparsification (MS) algorithm outperforms conventional techniques, effectively diminishing compression errors while maintaining the same compression level. Extensive experiments on various large-scale DNN models and datasets also demonstrate the effectiveness and efficiency of MS.

\subsection{Security and Reliability}

While promising for future communication networks, machine learning presents significant security and reliability concerns. Potential risks include privacy issues related to data sharing and malicious attacks on AI-empowered devices. SL may solve the problem by splitting the communication networks into parts, allowing the server and clients to train part of it separately. Since the model structure and parameters do not need to be shared between the server and clients, their models are black-box to each other~\cite{wang2023privacy}. Then, the raw data stays on the client's devices, reducing the risk of data breaches during transmission. The activations sent from clients to the server contain transformed information that is more difficult to interpret than raw data, adding an extra layer of privacy. This enhancement improves model security.

In SL, a portion of the neural network is resided and trained at the client side, ensuring that raw data is secure from leaks. However, exchanging the activation and gradient between clients and the main server poses a lot of potential risks of privacy attack~\cite{abuadbba2020can}. Attackers can perform Model Inversion Attacks or Label Inference Attacks to reconstruct the raw personal data with minimum error, or steal the labels~\cite{erdougan2022unsplit}. For black-box models, attackers can perform inverse attacks by means of, for example, adversarial samples~\cite{lecuyer2019certified}.

The authors in~\cite{abuadbba2020can} investigated sequential/time-series data (medical ECG data). Experiment results show that there is a high similarity between the split layer activation and the raw data when authors observed the disparity between them. Utilizing Distance Correlation and Dynamic Time Warping to measure the similarity between the original samples and the reconstructed version, the results indicate the raw input and the intermediate representations have a very high correlation. With the assumption of prior knowledge of the network’s architecture on the client side, servers can reconstruct pixel-perfect copies of the original images by model inversion and stealing attacks~\cite{erdougan2022unsplit}. Additionally, attackers can carry out a label inference attack on clients that hold only one layer of the DNN to obtain the labels of raw data. These evaluations are the outward sign of high information leakage and can be vulnerable to attack to infer the sensitive data in SL systems.

In order to limit data reconstruction, the authors in~\cite{vepakomma2021nopeek} introduced an approach to minimize the distance correlation between the raw data and the intermediary representations to prevent reconstruction attacks and sensitive information leakage during the training process. The proposed NoPeek-Infer combines the distance correlation loss and the task’s loss. The distance correlation loss is minimized between raw input data and the output of any chosen layer whose outputs need to be communicated from the client to another untrusted client or untrusted server. NoPeek-Infer's loss weighting is optimized to guarantee the intermediate tensors from the split layer have minimal relevant information to recover the original data. Also, this method maintains reasonable classification accuracies of the network. 

On the other hand, in~\cite{titcombe2021practical}, Laplacian noise is employed to add into activations before transferring to the central server, finding that the method can significantly reduce attack efficacy at an acceptable accuracy trade-off on MNIST. In study~\cite{oh2024privacy}, the authors propose a novel privacy-preserving SL framework that injects Gaussian noise into smashed data and mixes randomly chosen patches of smashed data across clients, coined DP-CutMixSL. It is a differentially private (DP) mechanism that strengthens privacy protection against membership inference attacks during forward propagation of large-scale vision transformer model training. Its simulations show that DP-CutMixSL improves privacy protection against membership inference attacks, reconstruction attacks, and label inference attacks, while also improving accuracy compared to baselines.

\subsection{Application Scenarios}

Previous content explores the fundamental benefits of SL in terms of enhancing communication efficiency and strengthening security in distributed machine learning systems. Building on these foundational strengths, the following delve into specific application scenarios where SL has been successfully implemented. We explore how SL is being applied to optimize semantic/goal-oriented communications, Integrated Sensing and Communication systems, and wireless caching. Each of these scenarios highlights the adaptability and effectiveness of SL in addressing unique challenges within different technological contexts, demonstrating its potential to enhance performance, reduce latency, and ensure data privacy across a wide range of applications.

\subsubsection{Semantic/Goal-oriented Communications}

Semantic or goal-oriented communications prioritize transmitting the intended meaning or achieving specific objectives over simply transmitting raw data \cite{ZHAO2024107055,yang2023secure,11008535,zhu2026semantic}. In wireless communications, the use of SL allows for processing and interpreting raw data locally on client devices, then extracting essential semantic features before sending intermediate representations to the server. This approach leverages context and relevance to optimize communication efficiency. In~\cite{choi2024semantics}, the application of SL in semantic communication demonstrates its effectiveness in addressing the challenge of semantic misalignment across multiple neural transceivers in multi-user scenarios. By leveraging SL with layer freezing (SLF), the system can locally process and fine-tune neural network layers, ensuring that semantic representations are aligned even when source data or channel conditions vary between different users. This approach optimizes communication efficiency by minimizing the need for data exchange while maintaining robust performance in semantic interpretation. SLF effectively balances the trade-off between communication costs and computational latency, making it a promising solution for enhancing interoperability and resilience in semantic communication networks. Future research could focus on minimizing end-to-end latency by optimizing the number of frozen layers, aiming for a target reconstruction error or classification accuracy. Furthermore, it would be beneficial to explore SLF applications with more users and in diverse tasks beyond reconstruction and classification.

\subsubsection{Integrated Sensing and Communication}

ISAC systems combine data sensing and transmission capabilities, facilitating simultaneous environmental monitoring and data communication \cite{dai2025joint,zhu2025digital}. SL supports this integration by enabling the initial processing of sensor data on local devices, extracting crucial features to minimize the data volume that needs to be transmitted to the central server for further analysis. Besides, the ability to dynamically adjust the split point in the neural network based on real-time conditions optimizes the utilization of both sensing and communication resources, enhancing overall system efficiency. In study~\cite{dantas2023split}, the authors present a multi-modal sensing-aided ML strategy based on SL that can cope with deployment challenges in novel radio access network (RAN) architectures. Proper and efficient beam selection is crucial for fully harnessing the potential of mmWave communications. Traditionally, each candidate beam is evaluated using reference signals (beam sweeping). However, the exhaustive search method can be time-consuming with high signaling overhead. This research applies multi-modal sensing information to the beam selection task enabled by SL, which is applied to cope with the challenges of using sensing data sources maintained by different vendors in an open and disaggregated RAN. In the proposed approach, accuracy levels above 90\% can be achieved while overhead diminishes by 85\% or more. SL achieves comparable performance with centralized learning-based strategies, with the added advantage of accounting for privacy and data ownership issues. Future directions for this research include the use of the proposed approach in multi-user equipment scenarios and the evaluation of the data pre-processing and SL's response time for real-time decisions.

\subsubsection{Wireless Caching}

Wireless caching involves storing frequently accessed content closer to the end users, thereby reducing latency and alleviating network congestion. SL enhances this process by analyzing user behavior to predict which content will be frequently requested. The initial analysis can be performed locally on client devices, and the results can be transmitted to the server to refine update caching strategies. This approach ensures a more responsive and efficient network. The adoption of SL in~\cite{chawla2024beyond} enhances the efficiency and responsiveness of network systems by optimizing the caching process. By analyzing user behavior locally on client devices and predicting the most frequently requested content, SL allows for a more intelligent and adaptive caching strategy. The client devices process raw data, extract essential features, and transmit only the necessary information to the central server. This reduces the overall communication burden while ensuring that cached content is highly relevant and aligned with user needs. The evaluation shows significant improvements over baseline FL techniques: the proposed approach achieves a reduction in computation by 1623x for image classification and 23.9x for 3D segmentation on resource-constrained devices. Additionally, it reduces communication traffic between clients and the server by 3.92x for image classification and 1.3x for 3D segmentation, while improving accuracy by 35\% and 31\%, respectively. SL's ability to minimize data exchange while maintaining privacy and reducing latency makes it a promising approach for improving the performance of wireless caching systems, particularly in resource-constrained environments such as IoT networks.

\subsection{Summary and Insights}

SL offers significant advantages for 6G wireless communication systems, including resource efficiency, privacy, and scalability. It reduces energy consumption and communication overhead compared to traditional FL, but challenges remain. Communication overhead can be mitigated by reducing the frequency and volume of data exchange, using techniques such as selective transmission, quantization, and sparsification. Privacy risks, such as model inversion and label inference attacks, can be addressed through noise injection and differential privacy mechanisms. SL has shown promise in applications like semantic communication, ISAC, and wireless caching, optimizing performance while reducing latency and overhead. Future research should focus on SL's scalability, interoperability, and robustness, and expanding it to diverse tasks and larger networks, ensuring its potential is fully realized in 6G systems.

\section{Communication Systems for Split Learning}% Xu
\label{sec_for_SL}

\subsection{Architecture and Model}

\begin{figure}
    \centering % 表示居中
    \includegraphics[width=0.7\linewidth]{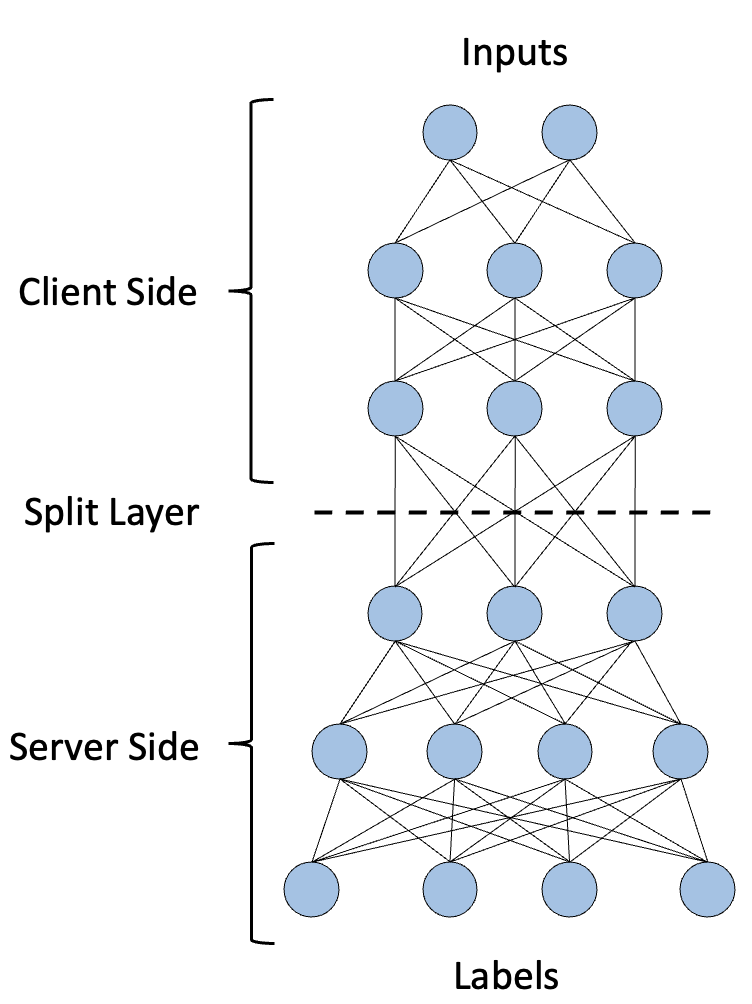}
    \caption{Overview of the SL mechanism.} 
    \label{fig_SL_mechanism}
\end{figure}

SL is a collaborative learning method. It eases the training overhead of clients by splitting the large model into parts: a smaller part held at the device side and the rest at the server, which allows clients and the server to train part of the model separately. The layer at which the model is split is called the ``split layer" as shown in Fig.~\ref{fig_SL_mechanism}. In the training process, the client performs the forward pass through its layers and sends the output tensor along with the labels to the server. The server continues the forward pass through its layers and computes the loss. It then starts the backward propagation of gradients through its layers and transmits the remaining gradients back to the client to continue the backward propagation through its layers. These steps are iterated until convergence is achieved.

\subsection{Resource Allocation}

\subsubsection{Key Performance Indicators}

The key performance indicators (KPIs) of SL, as mentioned in~\cite{zhou2023slperf}, are essential metrics to evaluate the performance and efficiency of SL systems. In summary, the primary KPIs for SL are:

\begin{itemize}
    \item \textbf{Efficiency} indicates how quickly an SL method can process and train on partitioned data. It can be quantified in terms of \textit{latency} (the time taken to process and transmit between the client and server) and \textit{throughput} (the amount of data processed per unit of time). These relate to computational costs and the speed at which the algorithm works, with lower latency and higher throughput contributing to faster model training and reduced resource consumption.
    \item \textbf{Communication Overhead} evaluates the amount of data that needs to be communicated between different nodes in the SL setup. It is typically assessed by the data transfer rate or the number of communication rounds required for model convergence. Lower communication overhead is desirable as it means less data needs to be transmitted, leading to faster training and less bandwidth usage.
    \item \textbf{Accuracy} of the SL models represent their ability to make correct predictions. This can be evaluated using standard metrics such as \textit{accuracy}, \textit{precision}, \textit{recall}, and \textit{F1-score}. The goal is to assess how well the SL technique performs compared to traditional centralized learning or other distributed learning techniques.
    \item \textbf{Robustness and Reproducibility} evaluate how resilient the SL method is. \textit{Robustness} refers to the SL system's ability to perform well under varying conditions, such as different client data distributions or adversarial attacks. \textit{Reproducibility} is concerned with the consistency of results when running the SL system multiple times with the same configuration, ensuring that results are reliable and stable over repeated experiments.
\end{itemize}

\subsubsection{Single-User SL Systems}

Single-user SL is a variant of the SL approach designed for scenarios where only one client (user) is involved. In single-user SL, the process involves only one client and one server. The client processes its data through the initial layers of the neural network and sends the intermediate representation to the server. The server then processes this representation through the remaining layers, computes the model's output, and sends back necessary information (such as gradients) for the client to update its part of the model.

The authors in~\cite{matsubara2020head} address the challenges of deploying DNNs on embedded systems, particularly in applications such as connected and autonomous vehicles (CAVs). The authors introduce a new concept called ``head network distillation" and ``bottleneck injection" to mitigate the limitations of traditional offloading solutions in vehicular edge computing. This technique involves designing student models that achieve in-network compression while minimizing computational load on mobile devices, without compromising accuracy. By using bottlenecks and quantization, they were able to significantly reduce communication delays in scenarios with limited wireless channel capacities. The evaluation shows that bottleneck quantization significantly reduces the file size of the bottleneck, as much as 75\% compression with respect to bottleneck output tensor and 86\% compression with respect to the resized input JPEG files, without impacting the accuracy.

In the internet of drones (IoD), where drones operate like IoT devices, the authors in~\cite{yao2023split} explore the challenges of conventional machine learning training methods. These methods require transmitting all data to a centralized base station, which raises privacy and security concerns. To address this issue, the authors introduce SL which divides the image classification machine learning model into a client-side model hosted by the drones and a server-side model located at the ground station. This approach ensures that data remains confidential and secure. The research focuses on exploring the practicality of integrating SL into IoD. The effectiveness of the proposed algorithm is assessed through simulations. The results show that the algorithm is highly effective in maintaining data privacy in IoD. With the growing demand for distributed machine learning systems that can train neural network models efficiently while preserving data privacy, the authors in~\cite{ayad2021improving} delve into the challenges of implementing SL. Recognizing the computation and communication overheads that can be prohibitive for edge devices in IoT systems with limited resources, the authors introduce a modified SL approach. Specifically, this novel system incorporates an autoencoder and an adaptive threshold mechanism to reduce both communication and computation overheads. When tested on an IoT system for autonomous driving applications, the modified approach showcased reduced overheads with only a minimal performance trade-off, thereby addressing the efficiency challenges of traditional SL in IoT applications. Regarding the diagnosis of COVID-19 using chest X-ray (CXR) images, the authors in~\cite{park2021federated} addressed the challenge of training neural networks using decentralized data while maintaining data privacy. The authors utilize FL, which shares neural network weights across clients, to address the issue of bandwidth-intensive weight exchange in complex network architectures. Their solution is to introduce ``split learning" to divide the neural network between client and server parts, thereby reducing computational and bandwidth demands. The authors' innovation lies in their use of the Vision Transformer~\cite{dosovitskiy2021image}, a new deep-learning architecture that can be easily decomposed and provides optimal performance in SL. Experimental results demonstrate that their approach achieves performance comparable to centralized training, even when dealing with non-identical data distribution simulating real-world hospital collaborations. Furthermore, when combined with multi-task clients, the approach may even improve diagnostic accuracy for COVID-19. 

The authors in~\cite{tran2022privacy} explore privacy-preserving learning models in the communication domain and introduce the concept of SL. They address concerns of data privacy in AI and machine learning by presenting SL as an alternative to traditional methods and provide comparative insights against other techniques such as FL and multi-party computation. The tutorial discusses the intricacies of SL, including its combination with other techniques, challenges in real-world implementation, and methods to enhance its efficiency and scalability. The paper showcases the practical applicability and efficacy of SL through various case studies, emphasizing its significance in ensuring data confidentiality in communication systems. Noticing growing concerns around data privacy in computer vision research, the authors in~\cite{wang2023privacy} focus on the challenges posed by using large-scale, cutting-edge models. To overcome the computational and communication obstacles these models present, the authors have developed a new lightweight framework called masked SL (MaskSL). This framework leverages the power of masking in SL, bolstered by the application of differential privacy to ensure strong security. The computational and communication overheads of various collaborative learning paradigms are thoroughly analyzed, demonstrating the superior performance of MaskSL. Empirical evaluations on real-world datasets, including applications such as face recognition and medical image classification, confirm the effectiveness and efficiency of the MaskSL approach.

For multimodal sensing data, the authors in~\cite{koda2020communication} present a new approach to predict mmWave received power for 5G wireless communication. To overcome the challenges of high communication overhead in conventional deep learning models, which require transmitting large datasets to centralized servers, the authors introduce a communication-efficient multimodal split learning framework. A key feature of this work is its emphasis on leveraging multimodal data. While previous research may rely on single data sources or types, this paper utilizes data from multiple sources or types to improve prediction accuracy. By incorporating multimodal data sourced from multiple sensors or types, the SL-based method can achieve better prediction accuracy.

Binarizing SL (B-SL) can reduce the risk of privacy breaches for SL-crunched data, while only slightly affecting model accuracy~\cite{dong2017dropping, yu2019distributed}. The authors in~\cite{pham2023binarizing} present a new approach to this process by binarizing the SL local layers. This method accelerates computations, resulting in up to a 17.5x reduction in processing time on mobile devices and up to a 32x savings in memory and bandwidth. It also significantly reduces the risk of privacy breaches. To further improve privacy protection, the authors suggested incorporating additional local leak loss training and differential privacy techniques. Visual reconstructions of the model from the split-layer activations of B-SL after optimization for leakage showed that some filters are set to zero to minimize leakage. Removing these filters with zero values could make the model lighter and reduce computational costs.

\subsubsection{Multi-User SL Systems}

As a collaborative learning framework, multi-user SL systems enable multiple devices to perform training tasks in parallel. The authors in~\cite{wu2023split} suggest a new SL approach to decrease training latency in SL by parallelizing the training process. In addition to dividing devices into clusters, multi-user SL systems facilitate collaboration among the entire AI model across clusters. Due to limited resources in mobile networks, the paper introduces a resource management algorithm to minimize the training latency considering device heterogeneity and network dynamics in wireless networks. By optimizing cut layer selection, device clustering, and radio spectrum allocation, the proposed solution can greatly reduce training latency compared to existing SL benchmarks while adapting to network dynamics.

In~\cite{tuli2022splitplace}, the authors introduce a new approach to efficiently running large-scale DNNs in MEC. This involves splitting and placing neural networks using both layer-wise and semantic-wise strategies. They propose an intelligent ``Splitplace" method that uses multi-armed bandits to address a split and placement problem. This method outputs layering and semantic splitting strategies aligned with service deadline demands. SplitPlace dynamically chooses between semantic and layer-wise splits to optimize both inference accuracy and the SLA violation rate for each incoming task.

To determine the optimal cut layer between devices and the server, the authors in~\cite{kim2023bargaining} introduce the bargaining game to achieve personalization and energy efficiency in multi-user SL systems. Specifically, the bargaining game is formulated as a multiplayer bargaining problem, where devices and the server negotiate to reach an agreement on the cut layer. The goal is to maximize the utilities of both devices and the server. The Kalai-Smorodinsky bargaining solution (KSBS) is obtained using the bisection method with the feasibility test to solve the problem. Simulation results demonstrate that the proposed framework achieves optimal sum utilities by balancing energy consumption, training time, and data privacy, and it is robust to non-iid datasets.

To minimize the effect of stragglers in multi-user SL systems, the authors in~\cite{samikwa2022ares} propose an adaptive resource-aware split learning framework by determining device-targeted split points based on time-varying network throughput and computing resources. For each training round, the ARES aims to minimize the round cost function by computing split vectors to reduce training time and system energy consumption. By addressing the challenges of resource constraints, heterogeneity of devices, and varying operational conditions in IoT environments, the experimental results demonstrate that the proposed scheme can accelerate model training by up to 48\% and reduce energy consumption by up to 61.4\% compared to FL and classic SL.

By utilizing asynchronous training, the authors in~\cite{chen2021communication} propose a training scheme for SL that aims to minimize communication overhead and computational requirements on the client side. The authors introduce the quantization of activations/gradients to 8-bit floating point prior to transmission, which further reduces communication overhead. Moreover, the client-side model is updated based on a loss threshold and processed in a round-robin fashion for each epoch. As a result, this scheme effectively addresses the issue of high communication overhead between edge devices and the server in SL. Experimental results on VGG11, VGG13, and ResNet18 models trained on the CIFAR-10 dataset demonstrate a significant reduction in communication cost (1.64x-106.7x) and client-side computations (2.86x-32.1x) with minimal accuracy degradation (less than 0.5\%) in the single-client case. The communication cost reduction is also observed for the 5 and 10 client cases (11.9x and 11.3x reduction on VGG11 models).

\paragraph{Over-the-Air Split Learning}

By utilizing wireless channels to realize fully connected layers through multiplexed signals, over-the-air split learning allows multiple computation nodes to collaboratively process machine learning tasks~\cite{yang2023over, yang2022over}. Therefore, over-the-air SL can provide an alternative to traditional NNs by leveraging the characteristics of wireless networks and optimizing the communication scheme design. Specifically, the authors in~\cite{yang2023over} propose an over-the-air split learning system consisting of three main steps: distribution, intermediate result transmission, and backward propagation. First, the ML model is divided into parts allocated to different computation nodes, and each node executes a portion of the model. Then, the intermediate results are transmitted from one node to the next, enabling distributed computation and communication. Finally, backward propagation is performed in reverse order, with each node executing the necessary computations. A theorem is proposed indicating that the introduction of Gaussian white noise in backward propagation does not affect the convergence rate of the SGD algorithm.

For multi-device edge split inference systems, the authors in~\cite{wen2023task} propose a task-oriented over-the-air computation scheme to directly maximize the derived discriminant gain and improve classification accuracy. To validate the performance of the proposed scheme, the authors design a human motion recognition system that allows multiple devices to perform a concrete inference task collaboratively. In this system, the authors apply the variable transformation to derive an equivalent difference of convex problem for joint steering power control and receive beamforming. 

\paragraph{Federated Split Learning}

Taking into account both the privacy-preserving aspects of FL and the efficiency enhancement of SL, federated split learning (FSL)~\cite{zhang2023split} is proposed as a solution for reducing training latency and achieving high accuracy in resource-constrained wireless networks.

In terms of efficiency and privacy, the authors in~\cite{turina2021federated} analyze the performance of FL and SL. Specifically, the authors demonstrate that FSL can reduce computation power consumption while maintaining high prediction accuracy with unbalanced datasets of paralleled clients. The NoPeek approach, which uses a loss function called distance correlation to measure the difference between the source data and the intermediate data, is leveraged to keep the source data private in FSL architectures. 

To speed up model training in resource-limited wireless networks, the authors in~\cite{zhang2023split} propose a group-based split federated learning (GSFL) framework to reduce communication overhead and computational load on mobile devices. Specifically, the GSFL framework consists of three major stages, i.e., the model distribution stage, the model training stage with forward and backward propagation, and the model aggregation stage. During the the model distribution and training stages, clients are partitioned into groups to reduce the need for data sharing and communication overhead for collaborative local AI model training. In addition, the authors employ split-then-federated learning for local training and a server-side model for edge server processing. Furthermore, the storage resources are optimized via a model distribution process while the model training efficiency can be improved. For UAVs in 6G networks, the authors in~\cite{liu2022wireless} introduce a multi-arm bandit (MAB)-based user equipment selection scheme to improve learning accuracy. They demonstrate that the proposed scheme can address the high communication overhead and privacy concerns under IID and non-IID data scenarios using a CNN model.

Against the backdrop of increasing data privacy concerns and the need for efficient distributed learning in multi-user FSL systems, the authors in~\cite{zhang2023privacy} address the trade-offs between privacy and efficiency by proposing the server weight update rules in FSL to minimize the memory usage on client devices. The authors first introduce the attack resilience metric as the misclassification rate of images caused by an attacker. The attack resilience metric is calculated by comparing the number of correctly classified reproduced images to the total number of reproduced images and thus can be used to evaluate the system's resilience to privacy inference attacks. Furthermore, the authors propose a client-based private approach via distance correlation inspired by the NoPeek, which leverages two loss functions to update weights. This approach employs an alternating loss function policy to enhance privacy by alternating between minimizing cross-entropy loss at the server and maximizing distance correlation loss at the client.

Considering the limited storage resources at client devices, the authors in~\cite{mu2023communication} tackle the challenge of reducing communication overhead and server storage requirements in multi-user FSL. FSL systems with multiple server-side models require substantial storage, impacting scalability and efficiency. In addition, these storage constraints can lead to a trade-off between model accuracy and the number of server-side replicas in FSL systems. Therefore, the authors utilize an auxiliary network to locally update client-side models, reducing the storage cost, which is proportional to the number of clients, for multiple server-side models. In this way, each client needs to keep only a single server-side model throughout the training process to minimize storage requirements.

Noticing the increasing computational demands and privacy concerns in multi-device collaborative training, the authors in~\cite{yin2023predictive} propose a predictive  generative adversarial network (GAN)-powered multi-objective optimization algorithm to balance training time and energy consumption. First, they design a parallel computing scheme for model splitting without label sharing and analyze the impact of delayed gradient on convergence. Then, they model the joint optimization of splitting decisions, bandwidth, and computing resources as a multi-objective problem. Instead of using a traditional Generative Multi-objective Optimization Evolutionary Algorithm, like NSGA-III, the authors utilize a GAN to predict solutions that dominate the current solutions, enhancing the search speed for promising candidates. The proposed GAN-powered algorithm first employs the discriminator to learn from the differences between dominating and dominated solutions using defined dominance pairs and then freezes the discriminator's parameters to train the generator, aiming to produce solutions that outperform the current dominating ones.

\subsubsection{Multi-Tier SL System}

In 6G wireless networks, multi-tier SL learning systems involve distributing DNN tasks across cloud, edge, and end devices to optimize inference and training for sensor fusion and fault tolerance~\cite{lin2023split}. Compared with multi-user SL systems, multi-tier SL systems can scale vertically by partitioning simpler NN layers on end devices and more complex layers in edge and cloud servers. Therefore, multi-tier SL systems not only enable local processing on end devices but also reduce communication costs and latency during end-edge-cloud collaboration. For instance, the authors in~\cite{Teerapittayanon2017DistributedDN} propose distributed DNNs across cloud, edge, and end devices for efficient inference and data privacy. During the evaluation of training, individual accuracies are defined for local, edge, and cloud models. In addition, the system utilizes entropy-based confidence criteria for early exit point selection during the inference processes. Furthermore, the softmax cross-entropy loss function is employed for optimizing the model performance during training. The evaluation is performed using a multi-view multi-camera dataset for object recognition accuracy improvement. The experimental results demonstrate that the proposed framework can find the optimal threshold for sample exit points to balance accuracy and communication costs.

Neural architecture search (NAS)~\cite{Zoph2016NeuralAS} is a technique for automating the design of artificial neural networks to optimize performance for specific tasks. It involves searching through a predefined space of possible network architectures to find the most effective configuration for end devices and edge servers with different resource constraints. By jointly considering the model split and NAS framework, the authors in~\cite{tian2022jmsnas} propose a gradient-based algorithm for non-differentiable latency metrics within the NAS process. In this framework, DNN compliance, splitting, deployment, and execution are integrated and can be formulated as a multi-objective optimization problem, including accuracy and latency. The learning process of DNNs in this framework can be formulated as computational graphs with latency and execution constraints. To optimize the performance, the authors propose a gradient-based algorithm for non-differentiable latency metrics within the NAS process, which demonstrates superior performance over state-of-the-art methods in large-scale image classification tasks.

In multi-tier SL systems, a large-scale of messages, which contain split-cost information crucial for determining the optimal split of ML tasks across the network, are transmitted to make split assignment decisions in a distributed ML system. To reduce communication overhead for message transmission, the authors in~\cite{wang2021hivemind} tackle the challenge of high information gathering overhead in distributed Dijkstra's algorithm for split ML problems by introducing split cost information (SCI) design, a distributed algorithm based on Dijkstra's logic, optimized for split ML graph representation. Specifically, the authors employ a graph representation to solve the multi-split problem efficiently. To minimize the inter-node signaling, the representation implements graph pruning and information aggregation and adopts a distributed algorithm to optimize message size and frequency, ensuring efficient updates. 

In integrated terrestrial and non-terrestrial networks (TNTNs), ubiquitous connectivity is established among ground nodes, satellites, unmanned aerial vehicles (UAVs), and so on. The authors in~\cite{wu2023split} introduce differential privacy-based defense mechanisms to counteract input reconstruction and label inference attacks that consider heterogeneous devices and networks. The DP-based mechanisms inject Laplace noise into intermediate data that disrupts the performance of potential privacy attacks. Furthermore, the authors discuss the digital twin-based network management of multi-tier SL systems, which involves real-time mapping of physical nodes to virtual counterparts for network management.

% \subsection{Application Scenarios}

% \subsection{Summary and Insights}

\section{Aggregation Learning for Communication Systems}% Mao & Sun
\label{sec_AL}

In 6G wireless communication, AL offers key advantages such as reduced communication overhead, improved privacy, and better scalability. It has already been widely applied in many distributed systems and holds significant potential.

\subsection{Architecture and Algorithms}

From a broad perspective, AL refers to any process within distributed learning that involves model aggregation. As a core step in distributed learning, model aggregation facilitates the integration of locally trained models from various nodes to construct a global model. Both SL and FL rely on AL to consolidate model updates from different devices, thereby optimizing the performance of the global model.

\begin{figure}
    \centering
    \includegraphics[width=1\linewidth]{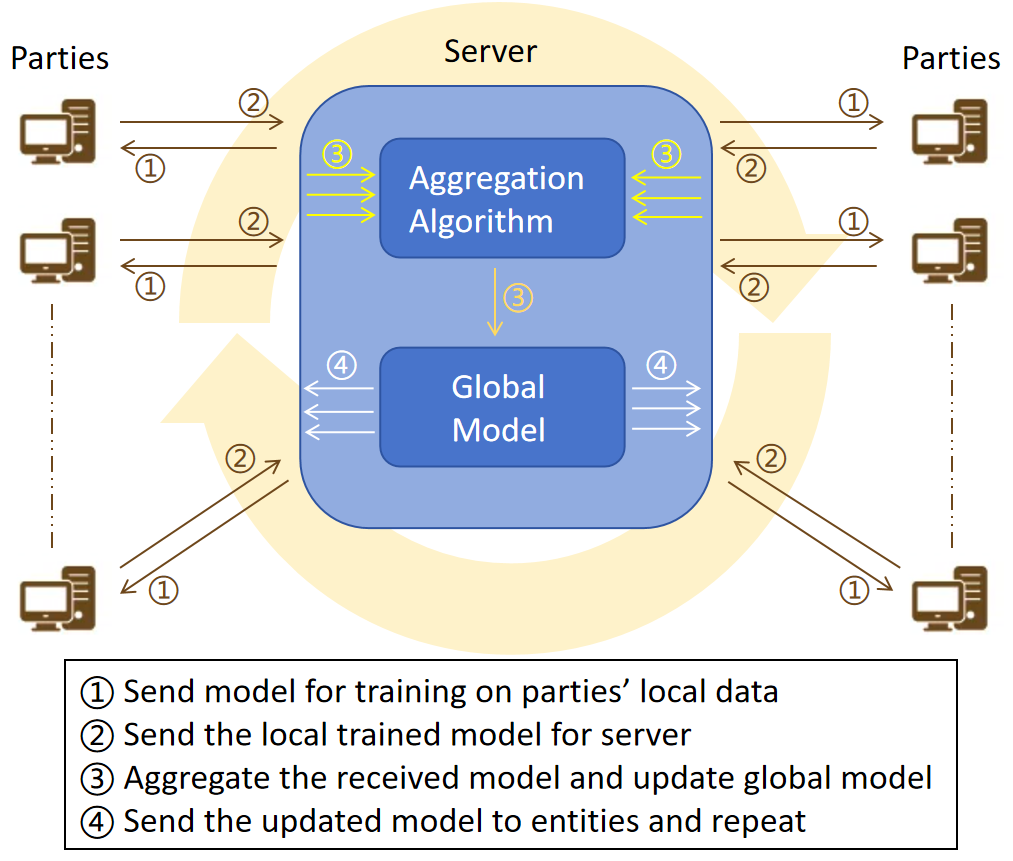}
    \caption{In a typical framework of FL, the architecture and process of AL can be observed. Parties train local models independently and send updates to a central server, which aggregates them into a global model for redistribution.}
    \label{fig_AL_architecture}
\end{figure}

Therefore, the AL architecture typically consists of multiple clients independently training their local models and sending the updated models to a central server. Upon receiving updates from clients, the server aggregates them to generate a global model, which is then sent back to the clients for further training, as shown in Fig.~\ref{fig_AL_architecture}~\cite{moshawrab2023reviewing}.

While the overall architecture of AL is well-defined, the specific aggregation strategies implemented at the server side can vary significantly in terms of communication efficiency, convergence behavior, and robustness. In the following, we introduce several representative aggregation algorithms—FedAvg, FedProx, and FedNova—that form the foundation for most AL implementations. These methods illustrate how local model updates are integrated under different assumptions and constraints, and serve as the computational basis for more advanced or specialized strategies.

\subsection{Communication Efficiency}
Numerous traditional aggregation algorithms have been proposed to address fundamental challenges, particularly those related to communication efficiency and overhead. These algorithms are commonly integrated into various federated frameworks as core computational approaches, yet they can also be applied to most other frameworks that involve model aggregation~\cite{qi2023model}.

\begin{figure}
    \centering
    \includegraphics[width=1\linewidth]{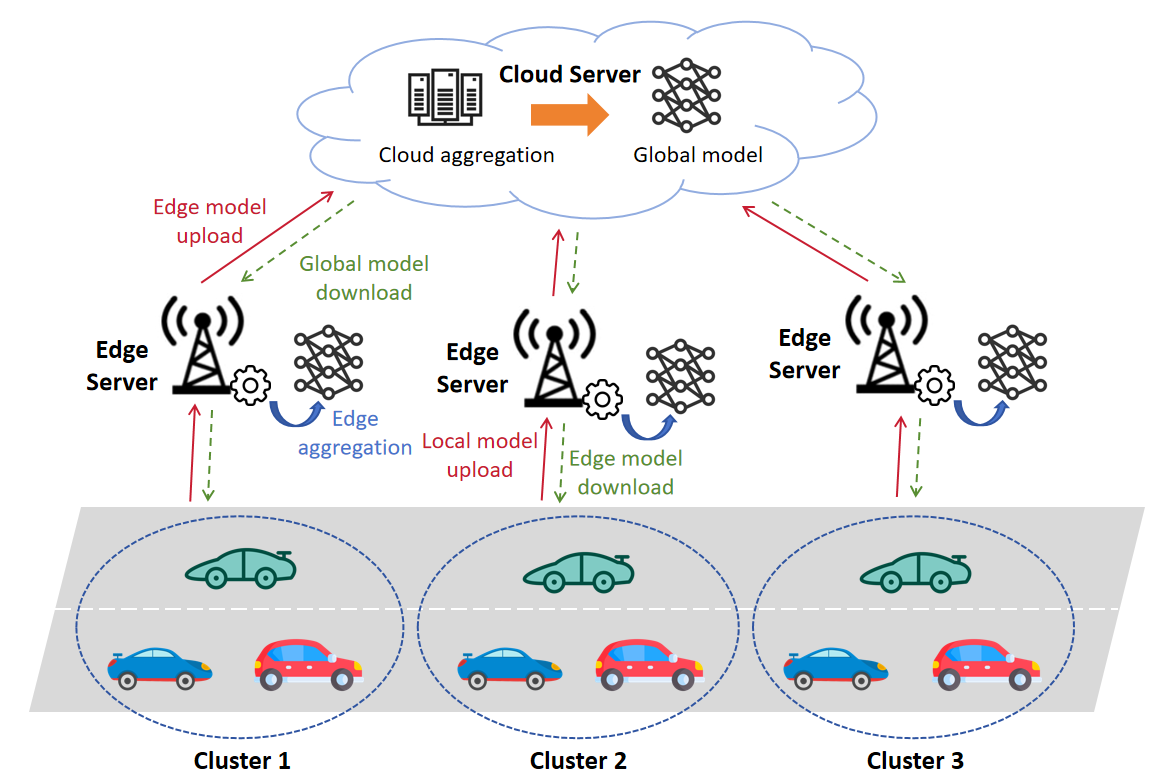}
    \caption{Hierarchical aggregation in FL. The figure depicts the hierarchical aggregation process through the federated IoV: vehicles are clustered into different clusters, and upload local models to the corresponding edge servers (i.e. roadside units), which then upload the models to the cloud for further global model updates.}
    \label{fig_AL_Hierarchical}
\end{figure}

The large number of edge devices, such as IoT devices, can significantly affect the efficiency of the learning process due to the frequent aggregation of models and the resulting high communication overhead~\cite{dinh2021enabling}. To tackle this issue, researchers have proposed a hierarchical aggregation approach that incorporates an edge layer, which performs partial aggregation of local models from closely related client devices before further aggregation occurs on the cloud server~\cite{luo2020hfel,tang2023research}. This hierarchical method aims to reduce both communication overhead and the number of model transfer rounds~\cite{deng2021share} by introducing multiple aggregation centers. Fig.~\ref{fig_AL_Hierarchical} illustrates the architecture of hierarchical aggregation, with hierarchy optimization and client similarity clustering being two key areas of focus in current research. 

In the context of the internet of vehicles (IoV), research~\cite{song2022federated,mirzaee2022chfl,lonare2021model} has explored the utilization of roadside units as the middle layer and the road traffic cloud as the cloud aggregator. This hierarchical approach leverages the infrastructure of the IoV to facilitate communication and coordination among vehicles. 

The concept of client clustering in Hierarchical Aggregation has gained significant attention in recent studies, as it presents an effective means to enhance communication efficiency and optimize resource allocation. Clustering involves categorizing clients based on shared characteristics or similar data distributions, enabling localized model aggregation within each group before performing broader aggregation at higher levels. Several clustering-driven techniques have been explored in existing research to reduce communication overhead and improve system scalability.

These methods are not directly applied to the core algorithm of AL, but rather enhance AL’s performance indirectly by improving communication mechanisms and optimizing data transmission strategies. For example, Lin \textit{et al.} proposed a method to group clients using device-to-device (D2D) communication, which reduces direct interactions with the central server. Within each cluster, D2D communication minimizes unnecessary data transmission, effectively alleviating network congestion issues ~\cite{lin2021semi}. By reducing communication overhead and optimizing data transmission strategies, this approach mitigates potential bottlenecks during AL model aggregation, thereby improving the overall efficiency of aggregation learning. Additionally, periodic global aggregation ensures the synchronization of model updates, allowing nodes to maintain consistency, which further enhances the effectiveness of the global model update.

Furthermore, the FedSim framework~\cite{palihawadana2022fedsim} applies the k-means clustering algorithm to assess client similarity and structure the aggregation process accordingly. This method uses different distance metrics, such as Manhattan distance, Euclidean distance, and cosine similarity, to identify the most efficient clustering approach, thereby optimizing data transmission and minimizing unnecessary overhead~\cite{wang2021adaptive,briggs2020federated,wang2022clustered}. By grouping clients and optimizing data transmission in this way, communication burdens during each model update can be significantly reduced, making AL training and aggregation more efficient in large-scale distributed environments.

% For instance, Lin et al.~\cite{lin2021semi} introduced a method where clients are grouped according to their communication capabilities, utilizing device-to-device (D2D) communication within each cluster to minimize direct server interactions. This setup reduces network congestion by limiting unnecessary transmissions, and periodic global aggregations ensure model updates are synchronized efficiently. The FedSim framework~\cite{palihawadana2022fedsim}, on the other hand, applies the k-means algorithm to measure client similarity and structure aggregation processes accordingly. Various distance metrics, such as Manhattan distance (L1), Euclidean distance (L2), and cosine similarity, are utilized to determine the most efficient clustering approach for minimizing data transmission overhead~\cite{briggs2020federated,wang2021adaptive,wang2022clustered}.

By incorporating client clustering strategies into hierarchical aggregation, communication overhead can be significantly reduced, as localized aggregations within clusters limit the frequency of global transmissions. Moreover, leveraging proximity-based clustering and adaptive similarity measurements ensures that model updates are more efficient, further improving communication efficiency across hierarchical learning systems.

\subsection{Privacy and Security}
In traditional artificial intelligence technologies, such as ML, the data used for training models is typically centralized in data centers. Once these data centers are attacked, the large volume of data that is leaked can result in incalculable losses. Compared to centralized ML, distributed learning methods offer stronger data privacy and security guarantees. However, recent research has identified several potential security vulnerabilities in models~\cite{isik2023arfed,liu2023privacy}. For example, attackers can infer participants' data distributions by analyzing the parameters of local models~\cite{mao2021romoa}, while malicious participants can influence the global model's security by introducing bad clients~\cite{yin2021survey}. These security risks have driven researchers to develop strategies to enhance the privacy and safety of aggregation learning methods~\cite{xu2021robust}.

In AL, model updates from multiple clients are aggregated into a global model, but the contributions from these clients may be inaccurate or malicious. Robust Aggregation is a method used to enhance the robustness of model aggregation in distributed learning, ensuring that even in the presence of malicious behavior or unreliable devices, the aggregation process remains stable and effective~\cite{qi2023model}. It can identify and filter abnormal updates from clients to ensure that the quality of the global model is not compromised. For example, techniques such as anomaly detection are used to identify and exclude malicious updates, ensuring they do not affect the aggregated model. Additionally, some systems use weighted aggregation, assigning lower weights to updates from unreliable clients, thereby ensuring that the final model is not skewed by malicious contributions~\cite{liu2023privacy}.  

Specifically, decentralized model aggregation is widely used in Robust Aggregation. Traditional centralized methods rely on a single server, which can lead to performance bottlenecks and single points of failure. To reduce this reliance, Gossip Principles have been proposed for decentralized aggregation. With this method, clients can efficiently exchange information without depending on a central server, thus reducing data transmission latency and enhancing the system’s robustness~\cite{belal2022pepper}. Additionally, this approach also strengthens data privacy, as the decentralized structure minimizes the risk of data leakage.

\begin{figure}
    \centering
    \includegraphics[width=1\linewidth]{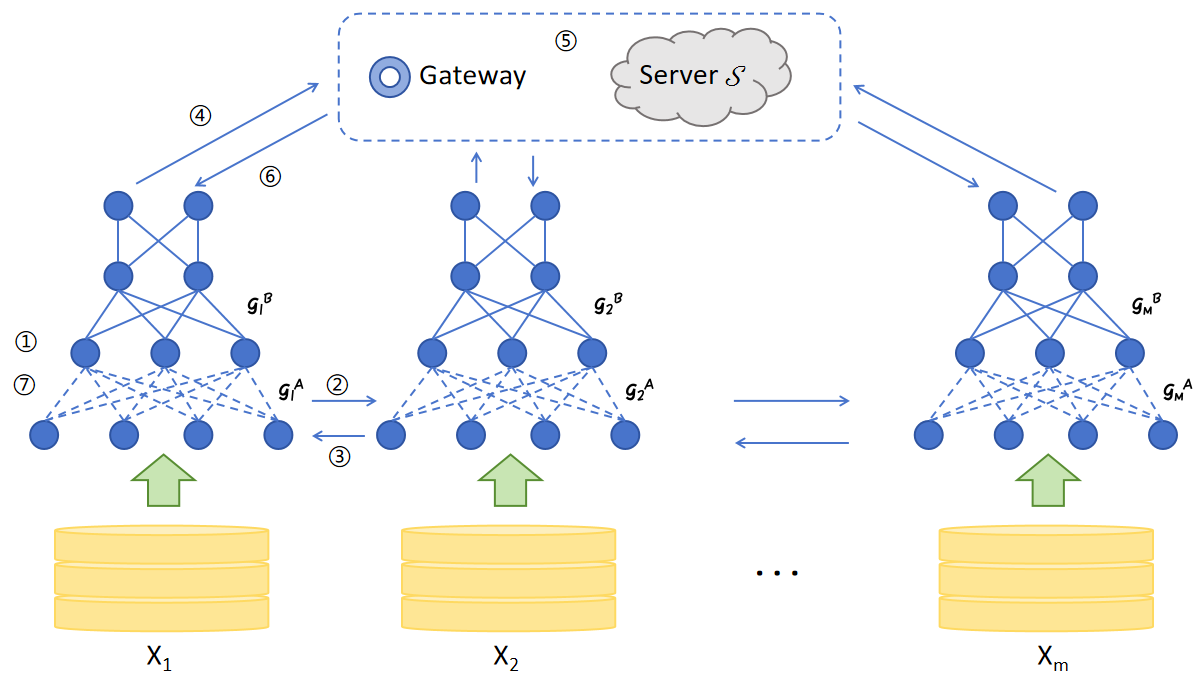}
    \caption{Framework for the partially encrypted MPC-based distributed machine learning. In each network, the Type A gradients are associated with the dash lines and the Type B gradients are associated with the solid lines. Only the Type A gradients will be aggregated using the secure MPC for encryption.}
    \label{fig_AL_MPC}
\end{figure}

Multi-party computation (MPC) is another important method to enhance aggregation robustness. MPC allows multiple parties to jointly compute a result without revealing each other’s data. This technology ensures privacy protection, particularly in scenarios involving sensitive data aggregation. For example, in distributed learning, clients use MPC to distribute their locally trained model updates to a selected group of users or servers. These users or servers then aggregate the updates, creating a new global model, while maintaining privacy~\cite{boer2020secure,sotthiwat2021partially}, as is as shown in Fig.~\ref{fig_AL_MPC}. This approach effectively prevents data leakage, as participants only receive the aggregated result and cannot view the specific data of other parties.
In summary, Robust Aggregation, through decentralization, MPC, and other privacy protection technologies, provides robust security for data aggregation in distributed learning. It effectively enhances the system’s robustness and privacy protection capabilities, making it particularly suitable for scenarios involving sensitive data aggregation in learning environments.

\subsection{Scalability}
As device heterogeneity becomes increasingly prominent in distributed systems, the asynchronous update feature of AL has garnered significant attention. Asynchronous aggregation allows clients to upload their local updates in a staggered manner, which helps mitigate the negative impacts of device heterogeneity. In traditional distributed learning frameworks, poor network signals or client crashes may lead to delays in uploading updates, thereby increasing the waiting time for the server to receive updates from all clients~\cite{zhou2021tea,chen2020communication}.

\begin{figure}
    \centering
    \includegraphics[width=1\linewidth]{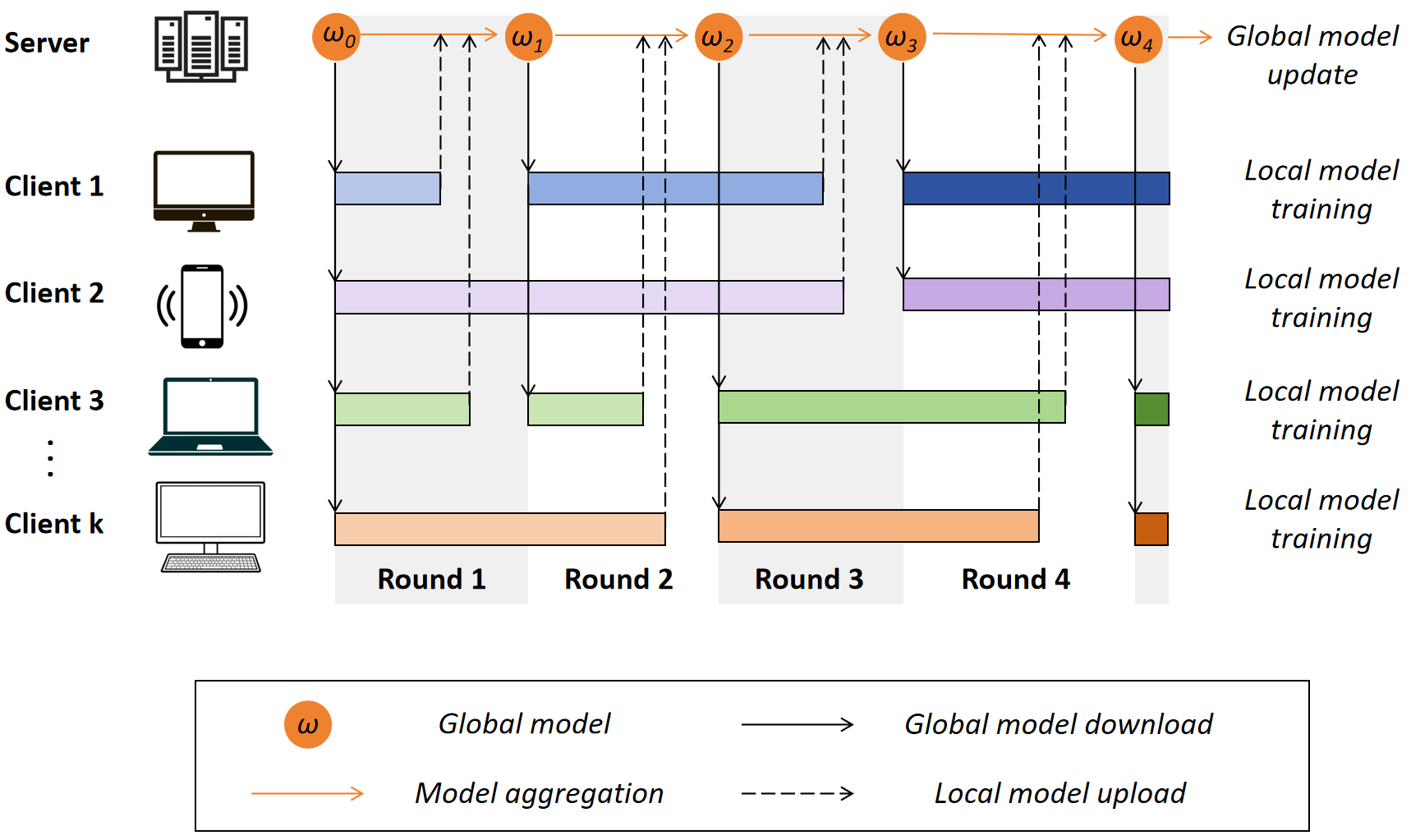}
    \caption{Asynchronous aggregation in an FL system. In the asynchronous aggregation process, the server carries out the aggregation in a set manner, as soon as updates from individual clients are received, without requiring synchronization with all clients.}
    \label{fig_AL_Asynchronous}
\end{figure}

The primary goal of asynchronous aggregation is to accelerate the training process. In fully asynchronous aggregation, once the server receives the local updates from a client, the aggregation operation takes place immediately. This enables each client to train independently without waiting for other clients to complete their updates. By reducing waiting times and allowing clients to submit updates at their own pace, this approach significantly enhances the efficiency and scalability of the system. A schematic diagram of asynchronous aggregation is shown in Fig.~\ref{fig_AL_Asynchronous}.

Semi-synchronous aggregation is a compromise approach that combines the characteristics of both synchronous and asynchronous aggregation~\cite{wei2022mssa}. In synchronous aggregation, the server must wait for all client model updates to arrive before performing the aggregation, which can result in unavoidable waiting times. On the other hand, fully asynchronous aggregation can effectively address the issue of device heterogeneity, but frequent model transmissions may consume significant communication bandwidth. To address these issues, researchers have proposed semi-synchronous aggregation learning mechanisms, such as FEDSA~\cite{ma2021fedsa}. FEDSA sets a communication budget that allows the server to perform partial aggregation based on the order in which client models arrive, thus finding a balance between optimizing aggregation waiting times and resource consumption. Semi-synchronous aggregation strikes a balance between synchronous and asynchronous aggregation, effectively improving the efficiency of the learning process while minimizing resource wastage.

% A comparison of the two aggregation methods can be seen in Fig. \ref{fig_AL_Comparison} . 
% \begin{figure}
%     \centering
%     \includegraphics[width=1\linewidth]{figures/image_23.png}
%     \caption{The comparison of workflows in asynchronous and synchronous aggregation in a FL system on heterogeneous devices.}
%     \label{fig_AL_Comparison}
% \end{figure}

In addition, recent research has proposed further improvements, such as dynamic fusion and evaluation of the client-side model, optionally for aggregation~\cite{qi2023model}. Through asynchronous and Semi-synchronous aggregation, AL exhibits good scalability in communication systems and holds potential for future updates and developments.

\subsection{Application Scenarios}

The unique advantages of AL have led to its widespread application in various communication systems such as financial services, healthcare, and intelligent transportation~\cite{qi2023flfd}. In these practical applications, selecting the appropriate aggregation method is crucial for ensuring system efficiency and privacy protection.

\subsubsection{Financial Services and Healthcare}
Despite the rapid development of ML, which has driven advancements in the intelligent healthcare and financial sectors, the issue of privacy data leakage poses significant challenges. To address this issue, aggregation learning can incorporate various privacy-enhancing techniques to build more secure systems~\cite{chetoui2023peer}.

% For example, a method proposed in~\cite{bonawitz2017practical} uses secure MPC, enabling the secure calculation of the sum of model parameter updates from individual user devices without exposing their updates. This method of computing multiparty sums, where no party reveals its update in the clear, is referred to as Secure Aggregation. Secure aggregation allows for the secure combination of local machine learning outputs from user devices to update the global model. Training models in this way offers significant advantages—when a user's device shares updates, the service provider only sees the information after it has been averaged with updates from other users, thus effectively protecting privacy.

For example, a method proposed in~\cite{bonawitz2017practical} uses secure MPC to securely compute the sum of model parameter updates from individual user devices without exposing their updates. This method is referred to as Secure Aggregation, and its core goal is to ensure the privacy and security of client updates during the aggregation process. In this approach, the model update results are only revealed after the updates from all devices have been securely aggregated, effectively protecting the privacy of user data. Secure aggregation allows the local machine learning outputs from user devices to be safely combined to update the global model. This method offers significant advantages — when a user’s device shares its updates, the service provider can only see the information after it has been averaged with the updates from other users, thus preventing the leakage of raw data.

Unlike Robust Aggregation, Secure Aggregation focuses on data privacy protection, ensuring that even the central server or aggregation node cannot access the detailed updates from individual devices. In contrast, Robust Aggregation focuses on ensuring the accuracy and stability of the global model by identifying and excluding inaccurate or malicious updates.

\begin{figure*}
    \centering
    \includegraphics[width=1\linewidth]{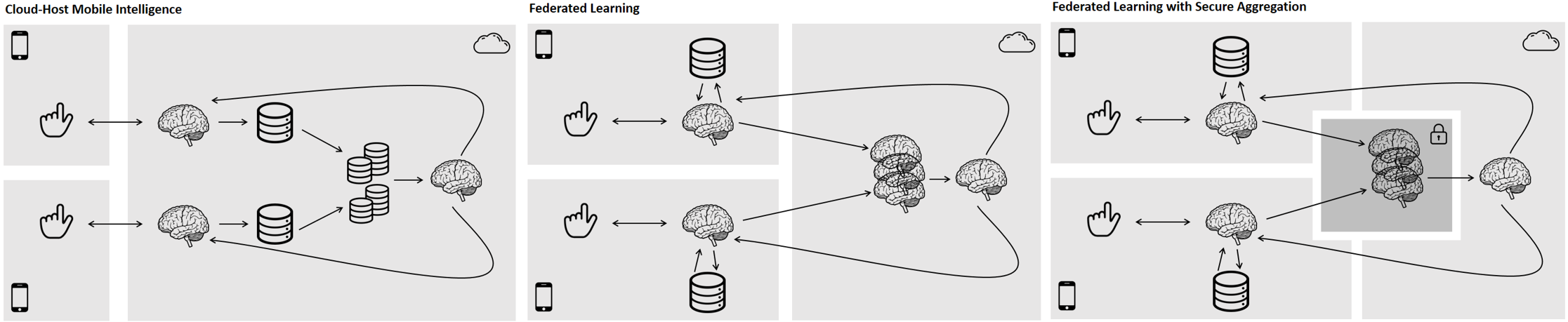}
    \caption{\fontsize{9.5pt}{11pt}\selectfont The comparison of Cloud-Hosted Mobile Intelligence, FL and FL with Secure Aggregation. Left: In a cloud-centric approach, user devices interact with cloud models, generating logs used for training. These logs are aggregated to improve the model, which is then deployed for future user requests. Middle: In Federated Learning, models are sent to user devices for local evaluation and training. Improved model summaries are shared with the server, aggregated, and redeployed to user devices. Right: With Secure Aggregation, model updates are aggregated by a secure, incorruptible third party using secure multiparty communication, ensuring the cloud provider only learns the aggregated update.}
    \label{fig_AL_Secure}
\end{figure*}

As is shown in Fig.~\ref{fig_AL_Secure}, when Secure Aggregation is added to Federated Learning, the aggregation of model updates is logically performed by the virtual, incorruptible third party induced by the secure multiparty communication, so that the cloud provider learns only the aggregated model update.

\subsubsection{Intelligent Transportation}

ML techniques are commonly used to enhance transportation, and researchers in the field of intelligent transportation typically employ various aggregation methods, considering the advantages each method offers~\cite{guo2023icmfed}. For instance, the FL-IoV framework based on hierarchical aggregation, proposed in~\cite{zhang2021distributed}, implements dynamic map fusion techniques without data labels. In this framework, roadside units provide local training labels, while the cloud server performs model aggregation. The use of vehicle edge computing in~\cite{ye2020federated} improves the accuracy and efficiency of model aggregation by employing a client selection method. Specifically, well-performing local DNN models are chosen based on evaluations of local image quality and computational capacity, then sent to the central server. In~\cite{liu2022fedcpf}, an asynchronous aggregation method is used for the local model aggregation of vehicles in IoV. Since a large number of vehicles upload model parameters to the server during the uplink communication phase in each round, this can cause significant communication overhead and extend training time. To address this, they implemented a framework with a participation rule for a subset of vehicles, allowing the system to converge faster with fewer communication rounds.

In addition, AL is widely applied in fields such as smart cities, industry, education, and network security, and is suitable for various distributed learning frameworks. In summary, due to the varying needs of different application areas and the distinct characteristics of communication systems, one or two primary aggregation methods are typically used to meet the specific requirements of each application.

\section{Communication Systems for Aggregation Learning}% Mao
\label{sec_for_AL}

As mentioned earlier, various AL methods are capable of adapting to communication environments with different characteristics and requirements. However, communication systems also play a crucial role in the development and application of AL. In distributed systems, communication systems typically refer to the networks and protocols that support efficient data transmission between different nodes, such as clients, edge servers, or data centers. These systems are responsible for ensuring the stability, reliability, and security of the information flow~\cite{hu2021distributed}.

 In recent years, the continuous development of efficient, stable, and secure communication systems has also advanced the application of AL, particularly in the context of distributed learning. Modern communication systems provide essential technical support for AL, including data encryption, secure model aggregation, low-latency communication, and large-scale parallel processing, ensuring both data privacy and system efficiency. These technologies lay the foundation for AL applications across various fields, particularly in scenarios like smart cities, autonomous driving, and healthcare, where efficient real-time computing and large-scale device collaboration are critical. The optimization of communication systems is especially vital in these contexts~\cite{hu2021distributed,tong2021federated}.

% AL is a distributed machine learning approach that focuses on integrating outputs from multiple models or features from various data sources to enhance decision-making, accuracy, and efficiency in complex systems. This method is particularly relevant in wireless communications, where it enables effective resource management, improves model robustness, and supports privacy-preserving computations across distributed networks\cite{mcmahan2017communication,niknam2020federated}.

\subsection{Resource Allocation}
Similar to SL, in AL systems, the three key performance indicators (KPIs) of the communication system are:

\begin{itemize}
    \item \textbf{Latency }refers to the time it takes for data to travel from one node (such as a client) to another node (such as a server). In communication systems, latency directly affects the speed of data transmission and system response times. Higher latency can cause delays in data transfer, which in turn impacts the timeliness and accuracy of the information, particularly in applications that require rapid feedback and decision-making, such as autonomous driving and smart traffic systems~\cite{liu2020efficient,konecn2016federated}.
\end{itemize}

\begin{itemize}
    \item \textbf{Scalability} refers to the ability of a communication system to maintain stable performance and effective operation as the number of devices or the amount of data increases. A highly scalable system can support the connection of more nodes and continue to operate efficiently and reliably even as the system load increases. In distributed learning environments, the increase in the number of devices requires the system to have good scalability to support parallel processing and large-scale collaboration~\cite{chen2018edge}.
\end{itemize}

\begin{itemize}
    \item \textbf{Reliability} refers to the ability of a communication system to operate continuously and reliably under various environmental conditions. A reliable communication system ensures that data is not lost during transmission and that the integrity and accuracy of the information are preserved. In distributed learning systems, the reliability of the communication system is crucial in preventing data loss due to network failures, signal interference, or device issues, ensuring that model training is not interrupted~\cite{hu2021distributed}.
\end{itemize}

Next, we provide a detailed discussion of how these three factors are manifested in different communication architectures and their impact on AL.

\subsection{Architecture and Model}

\begin{figure*}
    \centering
    \includegraphics[width=1\linewidth]{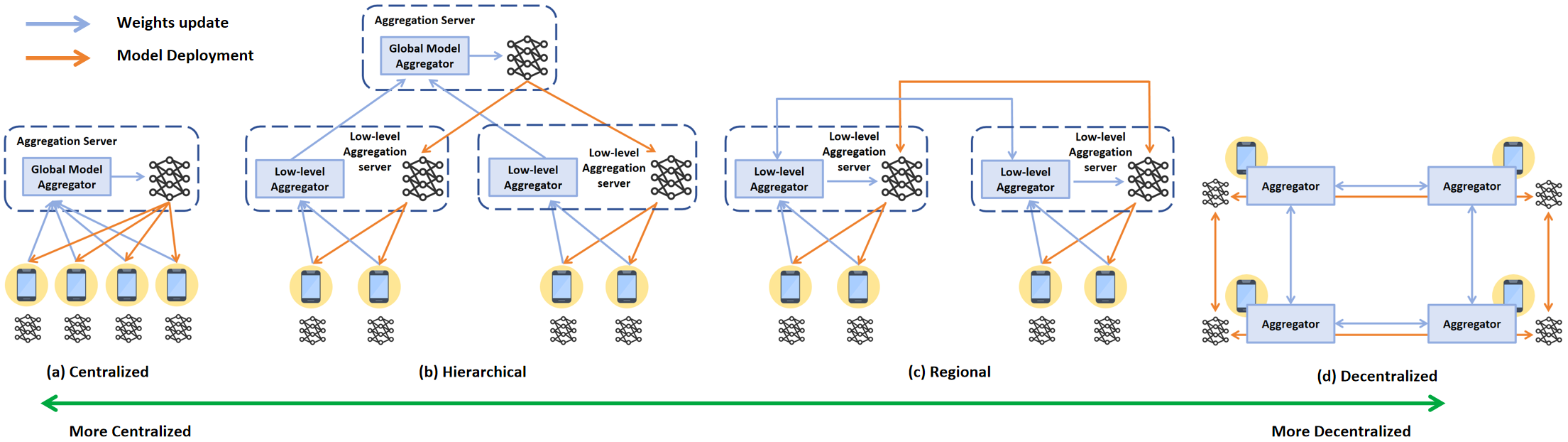}
    \caption{Architecture alternatives for distributed learning systems: centralized, hierarchical, regional and decentralized architecture. In the centralized architecture (a), all edge nodes connect to a central aggregation node for updating weights and distributing models. An improvement (b) introduces regional aggregation nodes to reduce data exchange and manage local devices. The regional architecture (c) removes the central management point to eliminate the single point of failure risk. The decentralized approach (d) moves aggregation entirely to the edge, allowing each node to perform local training and aggregation, ideal for situations where centralized servers face traffic bottlenecks. }
    \label{fig_AL_Architecture}
\end{figure*}

Centralized learning systems typically have several common architectures, the most common of which include centralized architecture, hierarchical architecture, regional architecture, and decentralized architecture, as is shown in Fig.~\ref{fig_AL_Architecture}. Each of these architectures has its advantages and disadvantages, and they require different types of communication systems to support their operation, as summarized in Table.~\ref{tab_AL_Architecture}.

\begin{table*}
    \centering
    \renewcommand{\arraystretch}{1.5}
    \caption{Comparison of Architecture Alternatives for Distributed Learning Systems.}
    \begin{tabular}{|>{\centering\arraybackslash}m{1.6cm}|>{\raggedright\arraybackslash}m{2cm}|>{\raggedright\arraybackslash}m{2.3cm}|>{\centering\arraybackslash}m{1.2cm}|>{\centering\arraybackslash}m{1.2cm}|>{\raggedright\arraybackslash}m{2.2cm}|>{\raggedright\arraybackslash}m{2cm}|>{\raggedright\arraybackslash}m{2cm}|}

\hline 
         \textbf{Architecture Type} & \multicolumn{1}{m{2cm}|}{\centering \textbf{Aggregation Method}} & \multicolumn{1}{m{2.3cm}|}{\centering \textbf{Communication Mode}} & \textbf{Scalability} & \textbf{Evolution Time} & \multicolumn{1}{m{2.2cm}|} {\centering \textbf{Applicable Scenarios}} & \multicolumn{1}{m{2cm}|} {\centering \textbf{ Pros.}} & \multicolumn{1}{m{2cm}|}{\centering \textbf{ Cons.}} \\
\hline
         \textbf{Centralized} & Single central node aggregates models & High traffic, all nodes communicate with the central server & Small scale & Slow & Medical applications, Human activity recognition & Easy to configure and manage, suitable for small systems & Poor scalability, communication bottleneck, high risk of single-point failure \\
\hline
         \textbf{Hierarchical} & Regional coordination nodes manage different edge clusters & Medium, regional nodes balance communication load & Medium scale & Slow & Mobile applications, Wireless systems & Load balancing, scalable for medium-sized systems & Still vulnerable to single-point failure, higher management cost \\
\hline
         \textbf{Regional} & Removes central aggregation node, multiple regional nodes aggregate models & Medium, local data exchange & Medium to large scale & Fast & Geographic location detection, Traffic applications & No central failure point, faster local model evolution & Higher hardware and management costs \\
\hline
         \textbf{Decentralized} & Each edge node aggregates locally & Low, local updates between nodes & Large scale & Very fast & IoT, Network-constrained systems & High autonomy, no bottlenecks, faster model evolution & Coordination issues, high management cost, difficult to obtain global knowledge \\
\hline
    \end{tabular}
    \label{tab_AL_Architecture}
\end{table*}

\subsubsection{Centralized Architecture}
According to the literature~\cite{hard2018federated,brisimi2018federated,lu2019collaborative}, centralized architecture is one of the most widely used architectures. In this setup, all communication operations are managed by a single central node. The central node communicates with all edge devices, aggregates local models, and deploys the global model. The central node is typically equipped with dedicated systems, which can be customized as needed. This architecture not only allows for rapid updates in small-scale systems but also makes it easy to disconnect any client node or server from the system without affecting other active nodes. For example, in medical applications and human activity recognition, centralized architecture is particularly suitable for small-scale systems, enabling quick deployment and updates, and it is easy to configure and manage\cite{zhang2020federated}.

However, centralized architecture also has certain limitations, especially in terms of scalability. As the number of client nodes increases, even with optimized hardware and software, the performance of the central server may still fail to meet the demands, particularly when thousands of devices are connected simultaneously. Additionally, when network traffic increases dramatically, communication bottlenecks may arise, leading to worsened data transmission delays. When the server is subjected to security threats such as distributed denial of service (DDoS) attacks, the system may quickly collapse~\cite{zhang2020federated}. These disadvantages of centralized architecture become especially apparent in large-scale distributed learning environments, particularly in wireless and mobile applications\cite{sun2021mind}. In these applications, thousands of devices may connect simultaneously, and the single central node can become a bottleneck, causing the system to operate inefficiently.

Therefore, while centralized architecture is easy to implement and update, its communication systems typically face issues of high latency, low scalability, and low stability. These limitations become particularly evident in large-scale distributed learning environments, especially when a large number of devices are connected. For example, in scenarios like weather prediction and traffic flow detection, centralized architecture may struggle to meet the demands of real-time data transmission and processing on a large scale\cite{zhang2020federated}.

In such environments, the communication system cannot support more efficient or flexible aggregation methods. These limitations make it difficult for the system to adopt more complex and resource-intensive aggregation strategies, such as asynchronous aggregation. As a result, synchronous aggregation becomes the most suitable choice. Model aggregation only occurs after all client updates have been received by the server, and this process typically causes slight delays or lags for the clients. Synchronous aggregation allows for the updating of the global model with relatively low communication overhead, and its structure is simple, making it easy to implement in a centralized architecture~\cite{qi2023model}. The advantages of synchronous aggregation are particularly noticeable in small-scale systems, especially in medical and human activity recognition systems, where low latency and high accuracy in real-time are critical\cite{hard2018federated}.

\subsubsection{Hierarchical Architecture and Regional Architecture }

Compared to the centralized architecture, hierarchical architecture introduces multiple regional coordination nodes to manage different edge clusters. These coordination nodes not only offload tasks such as model updates and aggregation from the central node~\cite{hard2018federated}, but also help distribute communication load and processing tasks, easing the burden on the central server. This architecture mitigates the communication bottleneck issues typically seen in centralized systems and performs well in medium-sized systems. By distributing tasks across regional nodes, the system can maintain relatively low latency and exhibit better scalability, making it suitable for environments with a growing number of devices and data. This architecture is especially advantageous in systems with a larger number of devices, where the task distribution across regions can avoid a bottleneck at a single node.

In mobile communication systems, for instance, as the number of devices and data traffic increases, hierarchical architecture improves system performance and responsiveness by distributing load across regional coordination nodes. By dynamically adjusting task distribution according to the network status in each region, it ensures stable system operation during peak traffic periods. This flexibility is crucial in managing mobile networks where the number of devices and data load can vary greatly\cite{zhang2020federated}. Hierarchical architecture shows its strengths particularly in medium-scale systems, where the number of devices is large but not yet at a scale that requires a fully decentralized system. In scenarios such as wireless communication or mobile applications, it balances load efficiently and ensures system stability. However, despite introducing regional nodes to reduce the central node's burden, hierarchical architecture still suffers from the risk of a single-point failure, particularly when facing network attacks or hardware failures\cite{derango2021novel}. 

Regional architecture, on the other hand, shares a similar design philosophy with hierarchical architecture but takes the additional step of removing the central aggregation node. Each edge cluster is assigned a regional aggregation node, which handles local model aggregation and updates~\cite{hu2018federated}. This decentralized design makes regional architecture more flexible and increases system resilience, as the aggregation process no longer relies on a central node. It enhances scalability and reliability, as each regional node independently performs aggregation and model exchange, reducing the dependency on the central node. This architecture also optimizes communication stability and maintains high performance in large-scale environments. The collaborative operation of regional nodes reduces the system’s reliance on a central server and further improves its robustness.

% For instance, in weather prediction and geographic location detection applications, regional architecture can localize computational tasks and data exchanges, significantly enhancing processing speed and model update efficiency. The aggregation of local data at regional nodes helps reduce communication delays, making it ideal for real-time systems like traffic monitoring and predictive maintenance in smart cities.

Furthermore, regional architecture excels in large-scale distributed systems. Since each regional aggregation node independently performs tasks, the system can avoid communication bottlenecks and ensure high reliability, even when handling large volumes of data. However, as the number of regional nodes grows, hardware costs and the complexity of managing server configurations also increase, especially in large-scale deployments. This represents a significant challenge in real-world applications, as maintaining system performance requires substantial resources\cite{zhang2020federated}.

Regional architecture is particularly well-suited for IoT and network-constrained systems, where a large number of edge devices need efficient local data processing and model updates. In these scenarios, regional nodes provide greater autonomy and reduce reliance on central servers, enhancing system flexibility and scalability. However, as the number of regional nodes increases, the cost of hardware and management also rises, making large-scale deployments more resource-intensive\cite{hu2018federated,qi2023model}. 

As a result, the communication systems in both hierarchical and regional architectures generally exhibit lower latency, higher scalability, and better reliability compared to the centralized system. These architectures, by distributing the load across multiple regional nodes, enable more efficient communication even as the number of devices and data volume increases. However, as the number of servers and regional nodes grows, the hardware costs and server configuration management challenges also increase, particularly as the system scales. This becomes a major challenge for real-world deployment as more resources are required to maintain the system's performance~\cite{zhang2020federated}.

Given the performance characteristics and structural design of these two architectures, both hierarchical architecture and regional architecture can effectively support hierarchical aggregation methods. While they are equipped with better hardware configurations and greater edge computing capacity than the centralized system, these architectures face challenges due to more frequent model updates and data exchanges. They cannot support excessively high transmission rounds or communication overhead. Thus, hierarchical aggregation becomes crucial. This method first performs aggregation of local models from closely related client devices at the edge layer, and then further aggregates at cloud servers or upper-level nodes. Hierarchical aggregation strikes a balance between communication efficiency and system load, optimizing resource allocation and enhancing overall performance~\cite{qi2023model}. Additionally, both architectures' communication systems can support semi-asynchronous aggregation, which allows aggregation to be performed based on the order in which client models arrive each round. This optimizes the balance between waiting time and resource consumption, providing a compromise that reduces communication overhead while maintaining efficient model updates.

\subsubsection{Decentralized Architecture}
Decentralized architecture is composed entirely of edge nodes. Unlike centralized or hierarchical architectures, decentralized architecture moves the aggregation function to the edge nodes for processing~\cite{hegedus2019gossip}. In this architecture, each node independently handles data processing and model updates, without relying on a central server or aggregation node. This design enables the system to effectively avoid the performance bottlenecks typically caused by centralized nodes, significantly improving system flexibility and responsiveness. However, despite the high degree of flexibility and autonomy offered by decentralized architecture, it still faces several challenges, particularly in terms of coordination. Since each node operates independently, coordinating global tasks and sharing global knowledge becomes complex and challenging, especially as the number of devices increases. The model differences between nodes can become significant, which may not always be optimal for certain scenarios.

For example, in intelligent transportation systems, each traffic light operates as an independent node, adjusting signal timings based on local traffic conditions. If each node makes decisions independently without relying on a central system, they can react more quickly at the local level, improving the system's responsiveness and efficiency. However, without coordination, conflicts between traffic signals can arise, affecting overall traffic flow\cite{qi2023model}. 

Despite these challenges, decentralized architecture, with its communication system characteristics of low latency, high scalability, and reliability, remains highly suitable for distributed learning applications that require high fault tolerance and flexibility, particularly in scenarios such as smart transportation, IoT, and large-scale data processing. In these applications, the efficiency and robustness of the communication system are key to successful implementation~\cite{zhang2020federated}. In IoT, millions of smart devices such as smart home devices, wearable devices, and sensors communicate using a decentralized architecture, with each device independently processing and updating its model. This decentralized approach allows for quick response when devices fail, reduces dependence on a central system, and minimizes bandwidth consumption\cite{yang2020federated}. 

In a decentralized architecture, the most suitable AL method is Asynchronous Aggregation. The communication system in this architecture supports frequent model updates, allowing each node to independently train and upload its model updates, which enables the system to handle more parallel computation tasks without imposing excessive computational or storage burdens on the central server. Since each node operates independently, the system can flexibly handle large-scale data exchanges and model updates without relying on central coordination, thereby improving overall efficiency. Furthermore, the communication system in decentralized architecture enables more efficient use of bandwidth. Since model updates primarily occur within local regions, there is no need to transmit all updates to a central server, which significantly reduces bandwidth consumption. This localized aggregation approach allows the system to efficiently exchange data and perform model updates even in high communication load environments, thus facilitating the smooth execution of Asynchronous Aggregation~\cite{wang2022asynchronous,xu2023asynchronous}.

\subsection{Summary and Insights}

As previously discussed, communication systems have always been a significant challenge for distributed model aggregation~\cite{yun2023slimfl}. For example, in common federated learning systems, multiple clients are often involved, such as in IoT-based federated systems that can comprise hundreds or thousands of devices. During model aggregation, a large number of clients need to upload their local updates to the same network, which can result in severe communication congestion due to limited network bandwidth~\cite{kairouz2021advances}. Although training relatively simple models can alleviate transmission burdens, the explosive growth of data in the internet era suggests that this approach may no longer be sufficient to meet the demands of real-world applications~\cite{hosseinalipour2022multi}. Therefore, maximizing the use of limited communication resources and improving aggregation efficiency become urgent issues that need to be addressed.

To tackle this challenge, various new aggregation schemes have been proposed to address communication bottlenecks. One such method is a fast aggregation approach in federated learning systems based on AirComp (Airborne Computation), which employs non-orthogonal multiple access (NOMA) technology~\cite{ni2022federated}. This technology utilizes the waveform superposition feature of multiple access channels to perform aggregate computations on data transmitted by multiple clients. Another proposed solution involves deploying multiple relays to assist signal transmission and improve the performance of airborne model aggregation, as suggested by Lin \textit{et al.}~\cite{lin2022relay,letaief2019roadmap}.

Furthermore, the emergence of 6G wireless communication is expected to provide an effective solution to communication bottlenecks. Compared to previous generations of wireless communication (4G and 5G), 6G offers higher data transfer rates, wider frequency bands, and broader network coverage~\cite{letaief2019roadmap}. The 6G era is anticipated to help distributed systems overcome the challenges they face in model aggregation, and these solutions could, in turn, promote the application of 6G aggregation methods across a broader range of IoT industries and AI services.

In summary, improving the efficiency and scalability of the aggregation process requires a combination of technological innovation and a deep understanding of the limitations of communication systems. In distributed learning, as the number of devices and data volume grows rapidly, relying solely on traditional communication architectures is no longer sufficient to meet performance requirements. Therefore, innovative communication technologies, aggregation algorithms, and system architectures must be leveraged to address these challenges. Technological innovation is not only reflected in the acceleration of model training and aggregation algorithms but also in optimizing communication system factors like bandwidth, latency, and stability, ensuring efficient data transmission and processing~\cite{yun2023slimfl}.

As the demand for more complex AI systems increases, ensuring the efficiency of large-scale communication will be crucial for the success of distributed learning applications. Particularly in scenarios such as smart cities, autonomous driving, and industrial IoT, real-time, efficient, and reliable communication systems are fundamental to supporting the collaboration of large-scale devices~\cite{qi2023model}. In this process, the advent of next-generation communication technologies like 6G is expected to provide the necessary network infrastructure for distributed learning, driving these technologies to be deployed and developed across a broader range of application fields. Therefore, how to balance technological innovation with the efficient utilization of communication resources will determine the success of distributed learning in practical applications.

\section{SL and AL with Emerging Communication Technologies and Applications}% Ji Thanks!
\label{sec_emerging}

\subsection{Joint Communication, Sensing, and Computation}
SL and AL can be effectively combined with joint communication and sensing systems to optimize resource utilization \cite{10353003,10375758}. For instance, in autonomous vehicles, SL enables the partitioning of deep learning models between the vehicle (sensing unit) and the edge server (computation unit). The vehicle processes raw sensor data up to a certain network layer and transmits the intermediate representations to the edge server for further processing. This approach reduces the communication burden and preserves data privacy, as raw data remains on the vehicle~\cite{tan2021integrated}. AL also plays a crucial role in intelligent driving systems. In research based on IoV, where a large number of vehicles are involved and complete coordination is required, hierarchical and asynchronous aggregation schemes are commonly used~\cite{zhang2021real}. In~\cite{liu2022fedcpf}, an asynchronous aggregation method is applied for local model aggregation of vehicles in IoV. Since, in each communication round, a large number of vehicles upload model parameters to the server during the uplink communication phase, this results in significant communication pressure and longer training times. To address this, their framework implements a participation rule for a subset of vehicles, enabling faster convergence in fewer communication rounds.

\subsection{Space-Air-Ground Integrated Network}
In SAGINs, SL can be utilized to distribute model training across different network segments. For example, satellites (space segment) can handle initial data processing, unmanned aerial vehicles (air segment) can perform intermediate computations, and ground stations (ground segment) can execute the final model training stages. This hierarchical SL approach leverages the unique capabilities of each segment, enhancing computational efficiency and reducing latency~\cite{meng2024semantics}. Moreover, by distributing the model training process across multiple network layers, AL can efficiently aggregate updates from various devices and nodes, ensuring the accuracy and consistency of the global model. For instance, local aggregation is first performed at the edge device layer, with partially aggregated results then passed up to higher layers for further integration. This multi-stage aggregation approach not only supports effective communication and data processing between different network segments (such as satellites, drones, and ground stations), but also significantly improves the overall system efficiency by reducing data transmission and optimizing computation processes~\cite{qi2023model,11006980}.

\subsection{Rate-Splitting Multiple Access}
Rate-splitting multiple access (RSMA) is a communication strategy that divides user messages into common and private parts, enabling efficient spectrum utilization~\cite{mao2022rate,10915662}. SL can be employed to process the common part of the message at the transmitter and the private part at the receiver, optimizing the encoding and decoding processes. And AL, through asynchronous aggregation of feedback from various receivers, ensures that information is effectively integrated during the model aggregation process. This information aggregation allows the base station to collect local updates (such as model weights or parameters) from multiple user devices and aggregate them into a global update, ensuring the overall efficiency of the system in a multi-user environment. Through the aggregation process of AL, once the updates from user devices are integrated, the base station can use this information to optimize subsequent encoding strategies, enhance spectrum utilization, and ensure that the communication needs of each user are adequately addressed.

\subsection{Quantum Communication}
In quantum communication networks, the distributed computation of quantum neural network (QNN) has been actively discussed for privacy-preserving information management due to the distribution of data and the model over multiple computing devices~\cite{qiao2025transition,zhao2024quantum}. In~\cite{park2023quantum}, the authors proposed quantum SL that splits a single QNN architecture between multiple distributed computing devices to avoid exposure of the entire QNN architecture. SL can be adapted between quantum and classical nodes, where the quantum node handles quantum data encoding and initial processing, and the classical node performs subsequent computations. This hybrid method leverages the strengths of both quantum and classical computing. The evaluation results verify that the proposed approach preserves privacy in classification tasks and also improves accuracy at most by 6.83\% compared to previous methods. 

\subsection{Sustainable and Scalable AI Systems}
% SL and AL contribute to the development of sustainable and scalable AI systems by reducing computational load and energy consumption. By partitioning models and aggregating updates, these approaches minimize redundant computations and data transmissions. For instance, in large-scale IoT deployments, devices can perform local computations and send only essential updates to a central server, where AL aggregates these updates to refine the global model. This strategy conserves energy and allows the system to scale efficiently as the number of devices increases~\cite{wu2022sustainable}.

SL and AL contribute to the development of sustainable and scalable AI systems by reducing computational load and energy consumption. By partitioning models and aggregating updates, these methods minimize redundant computations and data transmission~\cite{wu2022sustainable}. For example, in large-scale IoT deployments, devices can perform local computations and send only the necessary updates to the central server. These device updates are aggregated at the central server by AL, forming an ever-updating global model. This global model helps optimize multiple control strategies in the environment, such as improving energy efficiency, maintaining environmental stability, and enhancing collaboration among devices. Through the asynchronous aggregation process of AL, the system ensures consistency in real-time collaboration among multiple devices and flexibly improves and optimizes control strategies based on feedback from various devices. This approach not only reduces bandwidth consumption but also avoids over-reliance on the central server, allowing the system to scale efficiently and accommodate the addition of more devices~\cite{qi2023model}.

\subsection{Metaverse}
The Metaverse is a virtual shared space, which enables users to interact socially in a persistent online virtual environment. To generate high-level 3D environments, low-latency data transmission and learning-based sensor data analysis are required. With the advancement of 5G technology, both transmission delay and scene generation have improved in meta-applications. However, many Metaverse devices are battery-powered, and local processing and learning remain costly. In study~\cite{shu2024dynsplit}, the authors proposed a novel dynamic SL scheme for enabled Metaverse systems. In this scheme, each neural network is split into two segments, and the upper segment is stored at the base station side. Multiple pathways exist between two segments, each having distinct compression ratios and a gating mechanism that intelligently determines the selection of paths for each input data. This design excels at adapting to diverse Metaverse applications and network conditions, improving both the learning and computing phases of split models. The simulation results highlight the efficacy of our proposed scheme, showing that it does not hinder the convergence of split learning models. Future research could focus on optimizing communication and performance, potentially by integrating multiple dynamic networks and split points to find optimal solutions across a broader optimization landscape.

\subsection{Blockchain}
Integrating SL and AL with blockchain technology can enhance data security and trust in decentralized learning systems.~\cite{sai2024blockchain} proposed a Blockchain-enabled SL framework for collaborative learning in healthcare, with a novel client selection algorithm to select clients based on the data utility, the system utility, and the model utility. In this SL model, the neural network is trained collaboratively between the server and the clients, with the forward and backward propagation steps to update the weights. In the proposed framework, the Blockchain platform serves the functions of decentralized model governance, decentralized identity and access management, incentive management, and client selection governance. The experimental results indicate that the proposed SL model yields better results than the FL and cloud-centric machine learning models. In AL, blockchain technology has also been widely applied. Blockchain provides immutable records and trust mechanisms for decentralized learning systems, which are essential for ensuring data security and enhancing system transparency. Recent research~\cite{qi2022high} proposed a new method in which each aggregation node performs quality testing on its local model and broadcasts the reputation assessments to the blockchain network. Based on a combination of client contributions and reputations, the system allocates rewards to clients. This reputation-based reward distribution algorithm, coupled with blockchain technology, ensures quality assurance for model training~\cite{chen2021design,ranathunga2023blockchain}.

\subsection{AI Agents}
The rapid evolution of AI agents, autonomous entities capable of perceiving, reasoning, and acting within diverse environments, has been significantly propelled by advancements in multimodal capabilities \cite{zhao2026agentic,zhao2025agentic}. These capabilities enable AI agents to process and integrate various data types, such as text, images, audio, and video, facilitating more comprehensive understanding and interaction. In this context, SL and AL have emerged as pivotal frameworks for decentralized AI model training, offering solutions that enhance privacy, scalability, and communication efficiency. This survey provides an in-depth analysis of SL and AL within multi-agent systems, emphasizing their role in fostering collaboration among AI agents. By allowing each agent to train a segment of a shared model using local data and periodically share updates, SL and AL facilitate the aggregation of these updates into a global model that benefits all agents. For instance, in a fleet of autonomous drones, each drone can learn from its own experiences and contribute to a collective intelligence, thereby improving navigation and task coordination across the fleet~\cite{petrovic2018artificial}.
% NAS for network splitting optimization of SL: JMSNAS: joint model split and neural architecture search for learning over mobile edge networks

% SL with over-the-air computation: Over-the-air split machine learning in wireless mimo networks

% robustness? (unreliable notes, adversarial note(on purposely sending shit data or train an irrelevant part))

% Federated split learning

%\section{Challenges and Future Directions}
%\label{sec_challenges}

\section{Conclusions}
\label{sec_conclusions}

This paper has presented a comprehensive survey on SL and AL for wireless communication systems, focusing on their architectures, advantages, and integration with emerging technologies. SL and AL offer promising solutions for privacy-preserving, scalable, and communication-efficient AI training, enabling intelligent edge computing and decentralized model optimization. Their applications span a wide range of 6G-driven technologies, including semantic communication, SAGIN, RIS, and quantum communication. Despite their advantages, several challenges remain, including communication overhead, model synchronization issues, privacy risks, and the need for energy-efficient AI systems. Future research should focus on enhancing SL and AL architectures with adaptive aggregation strategies, improving security with cryptographic techniques, and optimizing resource allocation for large-scale deployments. Furthermore, as 6G networks evolve, SL and AL will play a critical role in self-learning AI-driven communication systems, enabling real-time, secure, and scalable AI solutions for next-generation networks. By integrating SL and AL with future wireless paradigms, this research provides insights into the potential of decentralized AI in transforming next-generation intelligent networks, paving the way for the future of AI-native wireless communications.

\bibliographystyle{IEEEtran}

\bibliography{main}
\end{document}